\documentclass[a4paper,12pt]{article}
\usepackage[left=2.4cm,right=2.4cm,top=2.0cm,bottom=2.2cm]{geometry}
\usepackage{comment}
\usepackage{enumitem}
\usepackage{color}
\usepackage[square,numbers,sort&compress]{natbib}
\usepackage{amsmath, amsthm, amsfonts, amssymb}
\usepackage{kotex}
\usepackage{graphicx}
\usepackage{adjustbox}
\usepackage{subfig}
\usepackage{mathtools}
\usepackage{chngcntr}
\usepackage{hyperref}
\usepackage{relsize}
\usepackage[T1]{fontenc}
\usepackage{xcolor}
\usepackage[framemethod=tikz]{mdframed}
\definecolor{grisframe}{gray}{0.97}
\definecolor{gristitleframe}{gray}{0.85}

\newtheorem{defin}{Definition}
\newtheorem{*defin}[defin]{$^{\spadesuit}$ Definition}
\newtheorem{thm}[defin]{Theorem}
\newtheorem{*thm}[defin]{$^{\spadesuit}$ Theorem}

\newtheorem{*lem}[defin]{$^{\spadesuit}$ Lemma}
\newtheorem*{lem*}{Lemma}

\newtheorem{*cor}[defin]{$^{\spadesuit}$ Corollary}

\newtheorem{*pos}[defin]{$^{\spadesuit}$ Postulate}

\newtheorem{*prop}[defin]{$^{\spadesuit}$ Proposition}

\newtheorem{*ex}[defin]{$^{\spadesuit}$ Example}

\newtheorem{*prob}[defin]{$^{\spadesuit}$ Problem}

\newtheorem{innercustomlem}{Lemma}
\newenvironment{customlem}[1]
  {\renewcommand\theinnercustomlem{#1}\innercustomlem}
  {\endinnercustomlem}

\newmdtheoremenv{thmf}{Theorem}

\def\RR{\mathbb{R}} 				% Number systems.
\def\ZZ{\mathbb{Z}}

\def\dd{\mathrm{d}} 				% Differential operator.
\def\pd{\partial} 					% Partial differential operator.
\newcommand{\norm}[1]{\left\Vert#1\right\Vert} 			% Norm.
\newcommand{\bbr}[1]{\left[#1\right]} 					% Big bracket - []
\newcommand{\sbr}[1]{\left(#1\right)} 					% Small bracket - ()
\newcommand{\at}[1]{\left.#1\right\vert} 				% Only right vertical line.
\newcommand{\bb}[1]{\mathbf#1} 							% Vector, operator boldfont.
\newcommand{\name}[1]{\textsc{(#1)}}					% Theorem names.
\newcommand{\ovl}[1]{\overline{#1}}
\newcommand{\twist}[2]{\mathbf{#1}\mathbf{#2}^\top-\mathbf{#2}\mathbf{#1}^\top}

\DeclareMathOperator{\where}{where}

\DeclareMathOperator{\spn}{Span}

\allowdisplaybreaks

\title{\LARGE STDP-based Associative Memory Formation and Retrieval}
\author{Hong-Gyu Yoon\footnote{hkyoon@unist.ac.kr} \;\;and\; Pilwon Kim\footnote{pwkim@unist.ac.kr, \textit{corresponding author}}\smallskip  \\
        Department of Mathematical Sciences \\
        Ulsan National Institute of Science and Technology(UNIST) \\
        Ulsan Metropolitan City \\
        44919, Republic of Korea \\ }
\begin{document}

\maketitle

\normalsize
\setcounter{equation}{0}
%\tableofcontents
%\listoffigures
%\listoftables

\begin{abstract}
Spike-timing-dependent plasticity(STDP) is a biological process in which the precise order and timing of neuronal spikes affect the degree of synaptic modification. While there has been numerous research focusing on the role of STDP in neural coding, the functional implications of STDP at the macroscopic level in the brain have not been fully explored yet. In this work, we propose a neurodynamical model based on STDP that renders storage and retrieval of a group of associative memories. We showed that the function of STDP at the macroscopic level is to form a “memory plane” in the neural state space which dynamically encodes high dimensional data. We derived the analytic relation between the input, the memory plane, and the induced macroscopic neural oscillations around the memory plane. Such plane produces a limit cycle in reaction to a similar memory cue, which can be used for retrieval of the original input. 
\end{abstract}

\section*{Introduction}
%\addcontentsline{toc}{subsection}{\nameref{sec_S1}}

Spike-timing-dependent plasticity(STDP), as a synaptic modification rule according to the order of pre- and post-synaptic spiking within a critical time window, has been demonstrated in the nervous systems over a wide range of species from insects to humans.  
STDP is considered to be critical for understanding the cognitive mechanisms such as learning of temporal sequences \cite{blum1996model, rao2001spike}, 
formation of associative memory \cite{tsodyks2002spike, szatmary2010spike} and manipulation of existing memory \cite{han2009selective,ramirez2013creating,redondo2014bidirectional}.
Despite such progress and findings, the question still remains open as to how STDP affects the distributed process of information at the macroscopic level in the brain.

Modeling macroscopic brain activity with nonlinear dynamical systems facilitates understanding of brain functions \cite{kelso1995dynamic,globus1995postmodern, breakspear2017dynamic}. The hypothesis of storing memory in a form of an attractor of the dynamics is now accepted with substantial supporting evidence \cite{wills2005attractor,rolls2007attractor, tsodyks1999attractor, stringer2005self, renno2014signature, rolls2010attractor}. However, it is still unclear how specific trajectories of neural states could emerge through neural plasticity. 

In this work, we propose that a neurodynamical function of STDP is related to storage and retireval of associative memories at a  macroscopic scale. When the system is excited by a repeating sequence, STDP create a circular set of directed connections inducing neural oscillations in the neural state space. While the neural state space is extremely high dimensional, the osillations are confined in a two-dimensional plane which we call \emph{memory plane}.
Such memory plane can act as a generator of a limit cycle in reaction to an external input. That is, 
once the system converges under a sequential memory input and forms the corresponding memory plane, it produces a limit cycle in reaction to a similar memory cue, which can be used for retrieval of the original input. 
%the stored memories can be dynamically revived around the memory plane if perturbed by a similar stimulus. 
%While the Hopfield network and other variants have attempted to encode memory to a fixed point attractor, they have little in common with actual brain dynamics which is based on oscillatory synchronization.

%When excited by a repeating sequence, STDP can create a circular set of directed connections inducing neural oscillations. The limit cycles in the neural state space have been considered to be involved in many functions including long- and short-term memory and storage of multiple memory components \cite{tsanov2009long, lega2012human, singer1999neuronal, buzsaki2004neuronal, siegel2009phase}. %Recently, it has been proposed that STDP can store transient inputs as imaginary-coded memories which give rise to stable oscillatory trajectories in the neural networks \cite{susman2019stable}. The results imply that memory can be learned and retained in a stable manner despite synaptic fluctuations in the brain. 
The presence and the function of such planar memory structure in the neural state space have caught attention in \cite{susman2019stable}, where it has been proposed that STDP can store transient inputs as imaginary-coded memories.
In this work, we formalized the concept of the memory plane and the retrievability of neural states to analyze how data is effectively stored in the neural state space.  We derived the analytic relation between the input, the memory plane, and the induced macroscopic neural oscillations around the memory plane. This enables us to understand the functional role of STDP in terms of neurodynamical systems and view the macroscopic neural oscillations in the brain as circulations across the memory representations. The analytic results in this paper suggest an alternative method to store and retrieve high-dimensional and strongly associated data sets in analog devices. In the separate work \cite{yoon2021astdpbased}, we proposed a practical encoding algorithm based on the analysis done in this article to store associate image/text data sets into retrievable neural states.

\section*{Model Setups}

\subsection*{Firing-Rate Neural Network with STDP}

Our work follows the framework of standard firing-rate models \cite{dayan2003theoretical, susman2019stable}. We set the differential equation for the neural state as
\begin{equation}\label{firing_eq}
\dot{\bb{x}}  = -\bb{x} + \bb{W}\phi(\bb{x}) + \bb{b(t)},
\end{equation}
where $\bb{x} = \bbr{x_1\;\cdots\;\;x_N}^\top\in\RR^N$ is the state of $N$ neuronal nodes and $\bb{W}=(W_{ij})\in\RR^{N\times N}$ is a connectivity matrix  with $W_{ij}$ corresponding to the strength of synaptic connection from node $j$ to $i$. Here  $\phi$ is a regularizing transfer function and $\bb{b}(t)$ is a sensory memory input. 

The mechanism of STDP can be formulated as \cite{kempter1999hebbian} 
\begin{align}
\dot{W}_{ij}(t) & = -\gamma W_{ij}(t) + \rho\underbrace{\left(\int_{0}^{\infty}K(s)\phi(x_j(t-s))\phi(x_i(t))\;\dd s\right.}_{\text{pre- to post- firing}}\notag \\
& \quad\quad\quad\quad\quad\quad\quad\quad\quad\quad+ \underbrace{\left.\int_{0}^{\infty}K(-s)\phi(x_j(t))\phi(x_i(t-s))\;\dd s\right)}_{\text{post- to pre- firing}}, \label{STDP}
\end{align}
where $K$ is a temporal kernel. The parameters $\gamma$ and $\rho$ are the decaying rate of homeostatic plasticity and the learning rate, respectively.

For analytic simplicity, we use $\phi(\bb{x}) = \bb{x}$ and a Dirac-delta kernel $K(s)$ defined as
\begin{equation}\label{kernel}
K(s) :=\begin{dcases}
\delta(s-s_0) & s > 0 \\
-\delta(s+s_0) & s \le 0,
\end{dcases}
\end{equation}
with $s_0>0$. After simplifications, the main model becomes

\begin{equation}\label{system}
\boxed{
\;\begin{dcases}
\dot{\bb{x}} = -\bb{x} + \bb{W}\bb{x} + \bb{b(t)} \\
\dot{\bb{W}} = -\gamma \bb{W} + \rho\sbr{\bb{x}\bb{x}_\tau^\top-\bb{x}_\tau\bb{x}^\top\;}
\end{dcases}}
\end{equation}
where $\bb{x}_\tau = \bb{x}(t-\tau)$ stands for delayed synaptic response. More detailed derivation of the evolution rule for $\bb{W}$ can be found in Appendix A.

\subsection*{Storage and Retrieval Phases}

Let $\bb{m}_1,\dots,\bb{m}_n\in\RR^N$, and each $\bb{m}_i$ be memory representations which are encoded from some external sensory inputs and are to be stored in the system  (\ref{system}).
We assume in the storage phase that the input $\bb{b}(t)$ takes a form of sequential oscillatory drive 
\begin{align}\label{memory_input1}
\boxed{\;\bb{b}(t) = \sum_{i=1}^{n}\sin(\omega t-\xi_i)\bb{m}_i, \quad 0\le\xi_1<\cdots<\xi_n<\pi,\;}
\end{align}
where $\omega$ stands for the frequency of neural oscillations and $\xi_i$, $i=1,\dots,n$ stands for the sampling time for each representation. 
In the next section, we will show that the synaptic connectivity $\bb{W}(t)$ converges to a certain constant matrix $\bb{W}^*$ that reflects the informations of memory representations $\bb{m}_1,\dots,\bb{m}_n$.

\begin{figure}[ht!]
\centering
  \includegraphics[width=0.95\textwidth]{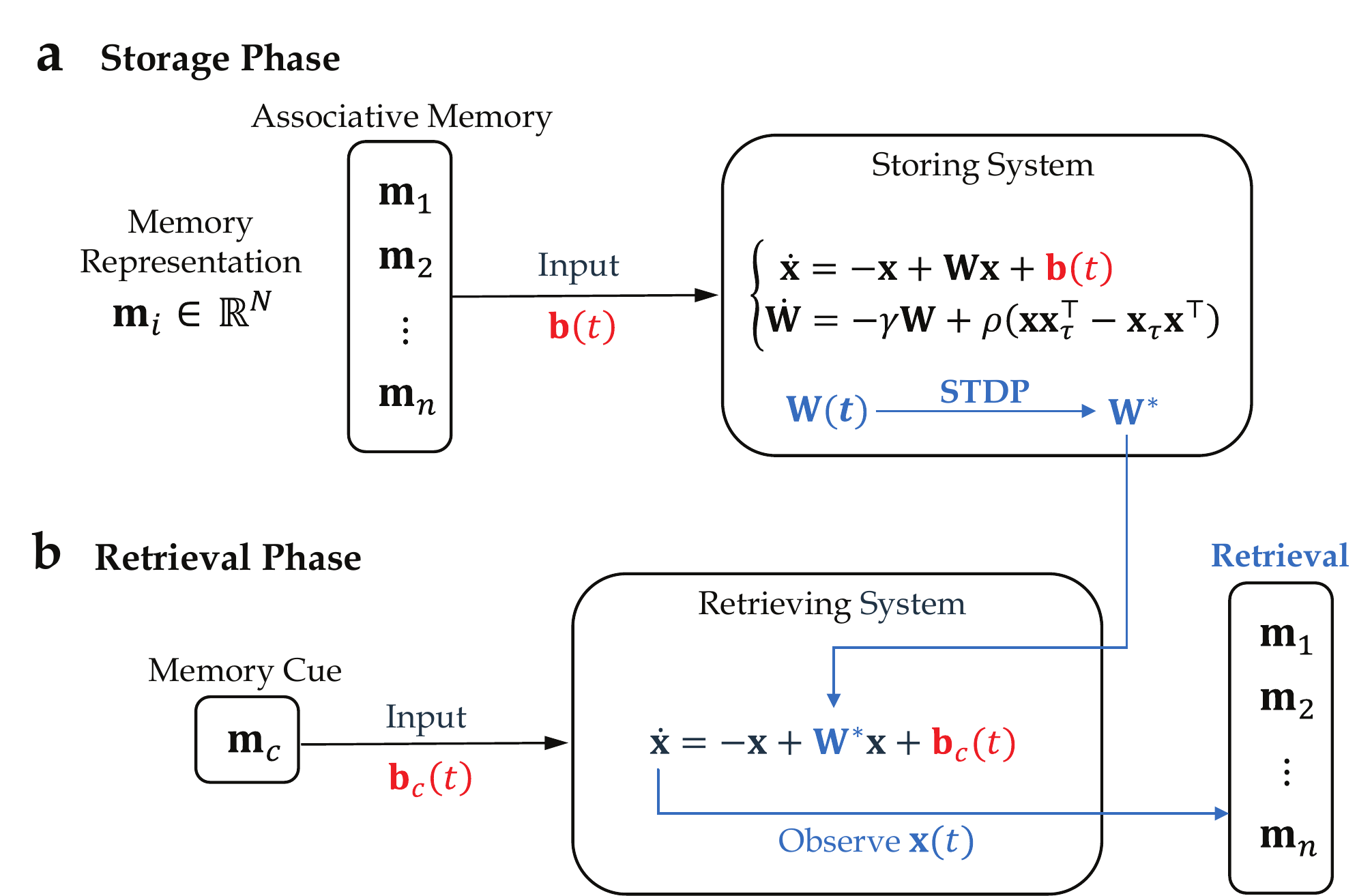}
  \caption{\small Description of the associative memory process for storage and retrieval of sensory input informations. \textbf{(a)} Storage phase: The STDP-based system processes the memory representations $\{\bb{m}_i\}_{i=1}^n$ and the connectivity matrix $\bb{W}(t)$ converges to a constant connectivity $\bb{W}^*$ as a result. \textbf{(b)} Retrieval phase: A memory cue input $\bb{b}_c(t)$ triggers the retrieval of the original inputs through the connectivity $\bb{W}^*$ acquired in the storage phase. } \label{fig_setup}
\end{figure}

In the \textit{retrieval phase}, change in synaptic weights is supressed (i.e., $\gamma=\rho=0$) as
\begin{align}\label{system_retrieval}
\boxed{\;
\dot{\bb{x}} = -\bb{x}+\bb{W}^*\bb{x}+\bb{b}_c(t),\;}
\end{align}
where  $\bb{b}_c(t)$ is the cue input in the form of 
\begin{align}\label{cue_input}
\boxed{\;
\bb{b}_c(t) = \sin\omega t\,\bb{m}_c, \quad \bb{m}_c\in\RR^N.\;}
\end{align} 
We are interested in how the original representations can be revived from the neural activity $\bb{x}(t)$ when $\bb{m}_c\in\RR^N$ is close to one of the memory representaions. Figure \ref{fig_setup}a and b illustrate the setup for storage and retireval process through the systems (\ref{system}) and (\ref{system_retrieval}), respectively.

\section*{Robust Learning by STDP}
This section presents some analytical results on the storage phase. We first confirm that
the sensory input in Eq. (\ref{memory_input1}) resides in a plane in $\RR^N$, a \emph{memory plane}, which is defined in the following lemma.

\begin{mdframed}[
backgroundcolor = grisframe,
innertopmargin = -4pt, 
outerlinewidth = 0.7pt,
leftmargin = 10pt, 
rightmargin = 10pt,
]
\begin{customlem}{A}\label{lem_plane}
$\bb{b}(t)$ is periodic and embedded in a plane $S:=\spn\{\bb{u},\bb{v}\}$ where
\begin{align}\label{uv}
\bb{u}=-\boldsymbol\Psi\sin\boldsymbol\xi\quad \mathrm{and}\quad \bb{v}=\boldsymbol\Psi\cos\boldsymbol\xi.
\end{align}
Here $\boldsymbol\Psi=\begin{bmatrix}
\bb{m}_1|\;\cdots\;|\bb{m}_n
\end{bmatrix}\in\RR^{N\times n}$, $\sin\boldsymbol\xi = \begin{bmatrix}
\sin\xi_1\;\cdots\;\sin\xi_n
\end{bmatrix}^\top\in\RR^n$, and $\cos\boldsymbol\xi = \begin{bmatrix}
\cos\xi_1\;\cdots\;\cos\xi_n
\end{bmatrix}^\top\in\RR^n$. 
\end{customlem}
\end{mdframed}

The following theorem asserts the existence of the periodic solutions $(\bb{x}^*(t),\bb{W}^*)$ of the system (\ref{system}) in terms of the memory plane $S$.

\begin{mdframed}[
backgroundcolor = grisframe,
outerlinewidth = 0.7pt,
innertopmargin = -4pt, 
leftmargin = 10pt, 
rightmargin = 10pt,
]
\begin{thm}\label{thm_1}
\name{Periodic Solution with Steady Connectivity}
The system (\ref{system}) under input (\ref{memory_input1}) has a periodic solution $\bb{x}^*(t)$ with a constant connectivity matrix $\bb{W}^*$, where
\begin{align}
\bb{x}^*(t)\in S\;\mathrm{for\;all}\;t,\quad\mathrm{and}\quad \bb{W}^* \in \wedge^2(S).
\end{align}
\end{thm}
\end{mdframed}

Here, $\wedge^2(S)$ indicates an exterior power of $S$, which is a set of anti-symmetric matrices in the form of $\alpha(\twist{\bb{v}}{\bb{u}})$ for any vectors $\bb{u}$ and $\bb{v}$ in $S$. The exact analytic form of such $(\bb{x}^*(t),\bb{W}^*)$ can be found in Appendix B2. Figure \ref{fig_storage} illustrates the convergence of the neural activity toward a periodic orbit $\bb{x}^*(t)$ on memory plane $S$ as described in Theorem \ref{thm_1}. Note that the memory plane $S$ does not necessarily contain the memory representations $\bb{m}_1,\dots,\bb{m}_n $ in general. However, we show in the next section that $S$ is likely located close to the memory representations in the high dimensional neural state space. 

%With the matter of its existence, the stability analysis in the next part approves that such solution pair $(\bb{x}^*(t),\bb{W}^*)$ is actually an attracting limit cycle for a certain parameter family of Eq. (\ref{system}). Further, the significance of this convergent connectivity $\bb{W}^*\in\wedge^2(S)$ will soon be revealed in a perspective of \textit{retrieval} in Section 4. 

%By these such importances of this two-dimensional plane $S$ on the evolutionary dynamics of STDP, we would like to use the term \textit{memory plane} for it throughout this paper. Fig. \ref{fig_storage} illustrates the convergence of the neural activity toward a periodic orbit $\bb{x}^*(t)$ on memory plane $S$.

\begin{figure}[ht!]
\centering
  \includegraphics[width=1\textwidth]{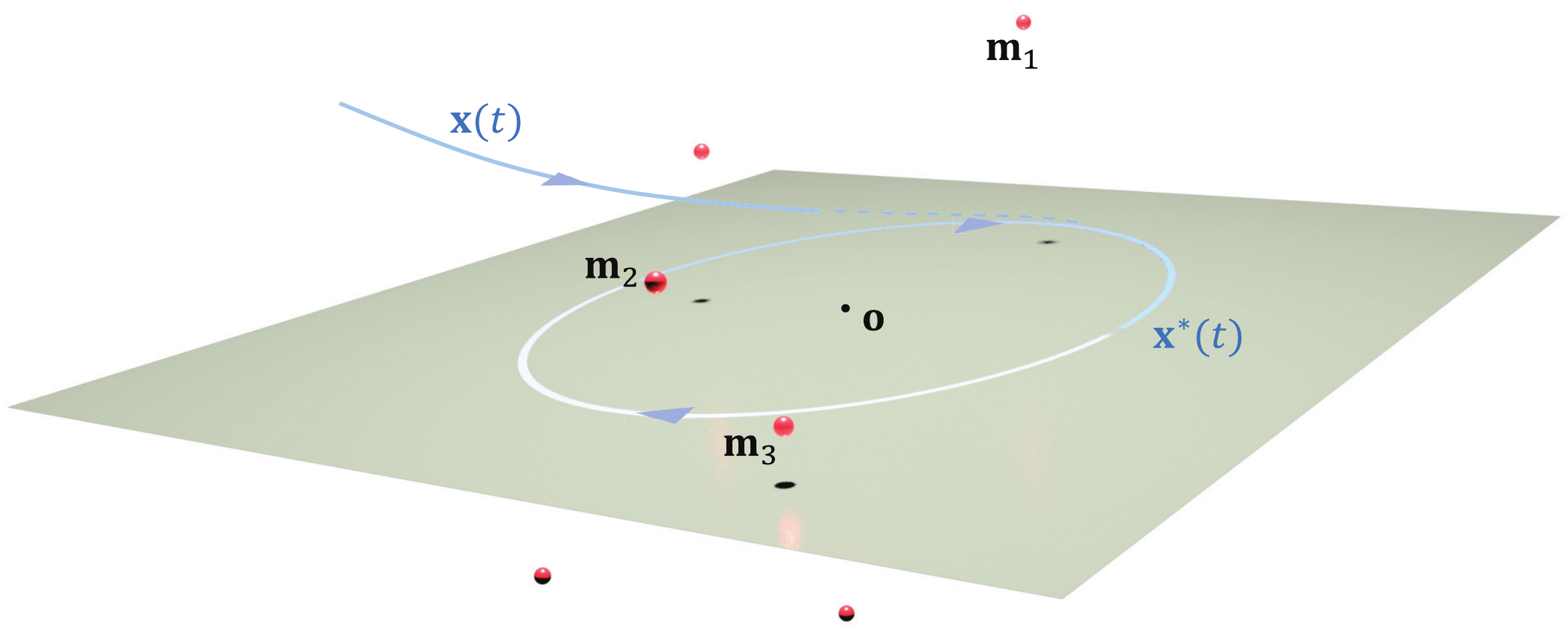}
  \caption{\small Illustrative image describing the convergence toward the periodic solution in Theorem 1. Each red circle represents the position of each memory representation $\bb{m}_i$ in the neural state space $\RR^N$. The memory plane $S$ is located close to the memory representations and plays a role of an attractor that brings $\bb{x}(t)$ to a periodic orbit $\bb{x}^*(t)$. } \label{fig_storage}
\end{figure}
To investigate the stability of $(\bb{x}^*(t),\bb{W}^*)$ found in Theorem \ref{thm_1}, we perform the analysis on the maximal Lyapunov exponent(MLE) \cite{sprott2003chaos, sandri1996numerical}. Setting $\bb{x}(t) = \bb{x}^*(t) + \delta\bb{x}(t)$ and $\bb{W}(t) = \bb{W}^*+\delta \bb{W}(t)$, we acquire  a variational equation from Eq. (\ref{system}) as
\begin{equation}\label{variational_dde}
\begin{dcases}
\dot{\delta\bb{x}} = (-\bb{I}+\bb{W}^*)\,\delta\bb{x} + \delta \bb{W}\bb{x}^* \\
\dot{\delta \bb{W}} = -\gamma\,\delta \bb{W} + \rho\sbr{\delta\bb{x}\,\bb{x}_\tau^{*\top}-\bb{x}_\tau^*\,\delta\bb{x}^\top + \bb{x}^*\,\delta\bb{x}_\tau^\top - \delta\bb{x}_\tau\,\bb{x}^{*\top}}.
\end{dcases}
\end{equation}
The derivation of Eq. (\ref{variational_dde}) and the detailed computational method for estimating MLE can be found in Appendix C1 and C2, respectively. Fig. \ref{fig_lyap} shows the color plot of numerically estimated MLE of Eq. (\ref{system}). For the regions showing negative values of MLE, one can assure that the solution $(\bb{x}^*(t),\bb{W}^*)$ is an attractor, thus consequently achieving a robust learning for any types of input of form Eq. (\ref{memory_input1}). 
\nopagebreak
\begin{figure}[ht!]
\centering
  \includegraphics[width=1\textwidth]{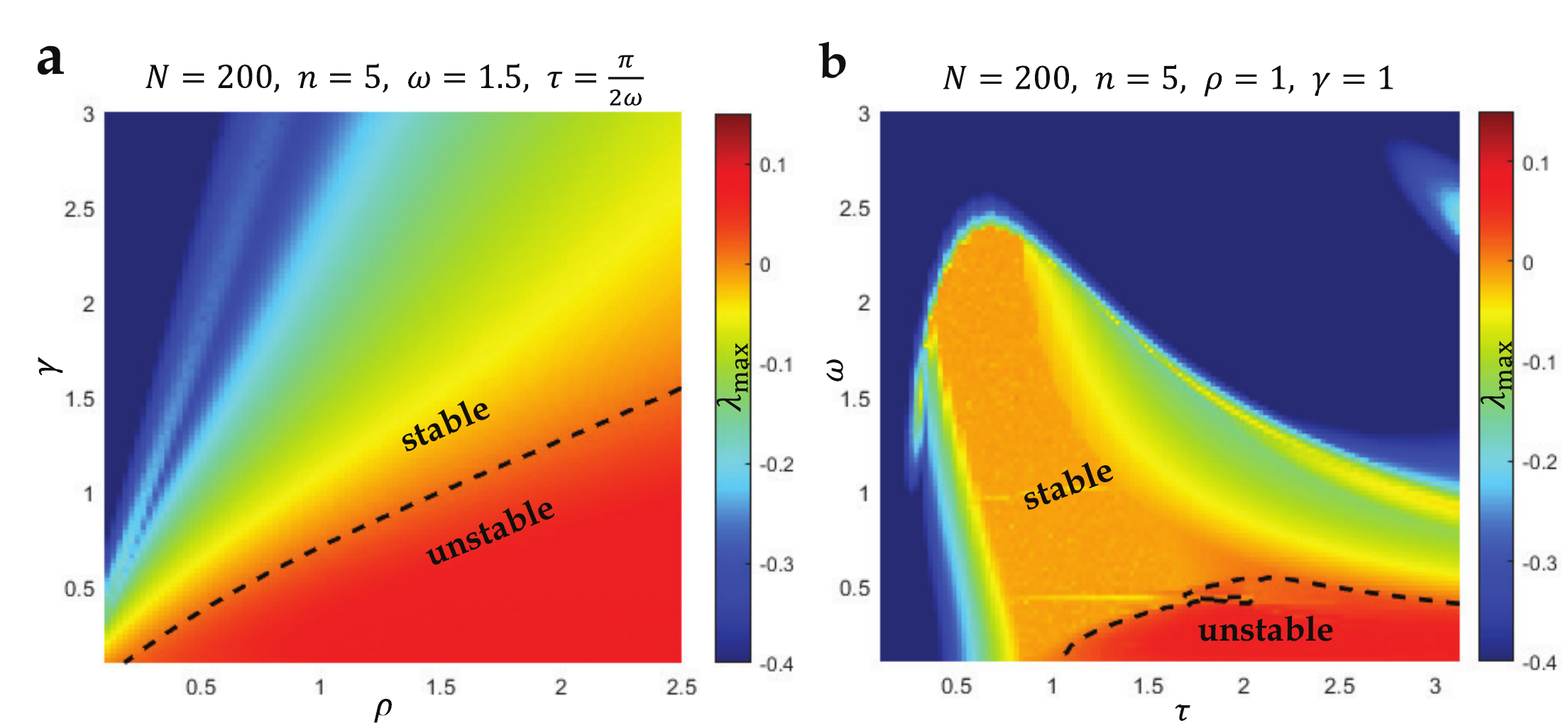}
  \caption{\small Plot of numerically estimated maximal Lyapunov exponent($\lambda_{\mathrm{max}}$) of Eq (\ref{system}), under input (\ref{memory_input1}) in the storage phase. \textbf{(a)} Color plot of $\lambda_{\text{max}}$ for parameter $(\rho,\gamma)\in[0.1,2.5]\times[0.1,3]$ for system with number of nodes $N=200$ under input of 5 unit length normalized memory representations and $\omega =1.5$, $\tau=\frac{\pi}{2\omega}$. The unstable region tends to be confined in $\gamma\le\alpha\rho$ with some $\alpha\approx 0.7$. \textbf{(b)} Plot of $\lambda_{\text{max}}$ for parameter $(\tau,\omega)\in[0.1,\pi]\times[0.1,3]$ for system with number of nodes $N=200$ under input of 5 unit length normalized memory representations and $\gamma=\rho =1$. For both plot, generally, the stale/unstable regions are hardly affected by the size of the system $N$. On the other hand, unstable regions tends to grow larger if the size of the input $\bb{b}(t)$ (or analogously, the number of memory representations $n$) increases.  } \label{fig_lyap}
\end{figure}

\section*{Auto-associative Retrieval by a Memory Cue}

In this section, we provide the analysis on Eq. (\ref{system_retrieval}) under cue input Eq. (\ref{cue_input}). We propose that the convergent synaptic connectivity $\bb{W}^*$ acquired from the storage phase effectively contains the information of a whole set of memory representations $\{\bb{m}_i\}_{i=1}^n$ and leads to periodic retrieval of them.

Let us define a \textbf{retrievable subspace} $\mathcal{M}:= \mathrm{Span}\{\bb{m}_i\}_{i=1}^n$ with respect to a set of memory representations $\{\bb{m}_i\}_{i=1}^n$. A neural state $\bb{x}\in\RR^N$ is said to be \textbf{retrievable} with respect to $\{\bb{m}_i\}_{i=1}^n$, if $\bb{x}(t)\in\mathcal{M}\setminus\{\bb{0}\}.$ Note that the memory plane $S$ is  a subset of the retrievable subspace $\mathcal{M}$ (see Eq. (\ref{uv})). 
In the separate work \cite{yoon2021astdpbased}, we work on a practical implementation of the system (\ref{system}) with some encoding/decoding processes, and show that a series of external sensory data can be recovered from a retrievable state $\bb{x}(t)$  
as long as they are properly encoded into the memory representations $\{\bb{m}_i\}_{i=1}^n$. Refer to Discussion section for more about decoding of retrievable states.

\begin{figure}[ht!]
\centering
  \includegraphics[width=0.55\textwidth]{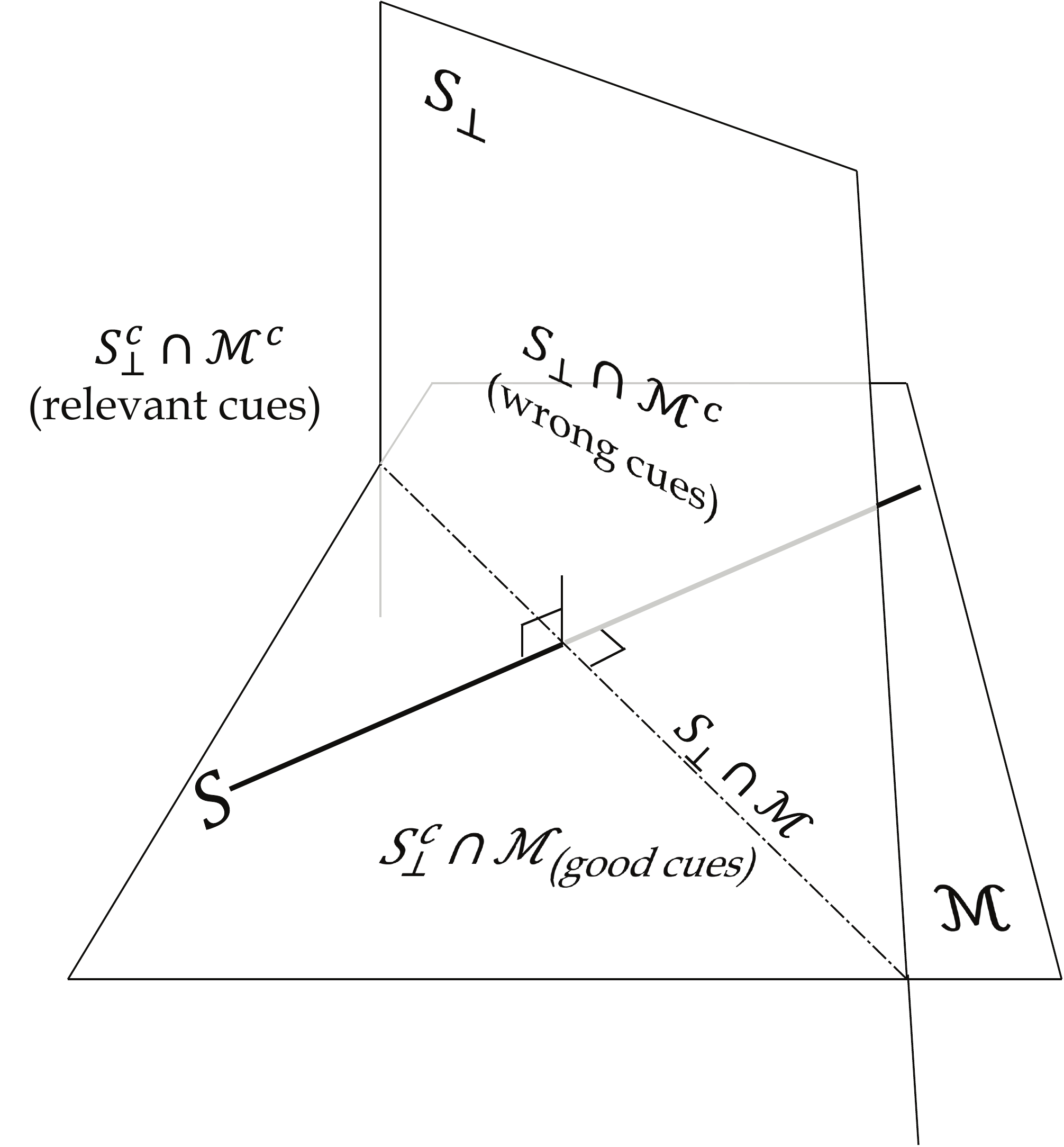}
  \caption[Dependence of the retrieval dynamics on the memory cue $\bb{m}_c$]{\small  For the intuitive graphical understanding, the retrieval subspace $\mathcal{M}$  and the memory plane $S$ are visualized as a plane and an embedded line, respectively.  (i) Case $\bb{m}_c\in\mathcal{M}$ (good cue): $\bb{x}^*_r(t)\in\mathcal{M}$ for all $t$.
 (ii) Case $\bb{m}_c\in S_\perp^c\cap\mathcal{M}^c$ (relevant cue): $\bb{x}^*_r(t)$ is retrievable at $t=t^\dagger$ as in Theorem 2.
 (iii) Case $\bb{m}_c\in S_\perp\cap\mathcal{M}^c$ (wrong cue): $\bb{x}_r^*(t)$ never becomes retrievable. } \label{fig_illust}
\end{figure}

The following theorem states that for some appropriately chosen memory cue representation $\bb{m}_c$,  there is a specific moment $t=t^\dagger$ at which the corresponding neural state $\bb{x}(t)$ becomes retrievable.
\begin{mdframed}[
backgroundcolor = grisframe,
outerlinewidth = 0.7pt,
innertopmargin = -4pt, 
leftmargin = 10pt, 
rightmargin = 10pt,
]
\begin{thm}\label{retrieval_thm} 
\name{Periodic Retrieval}
%Consider a set $\mbr{\bb{m}_1,\bb{m}_2,\dots,\bb{m}_n}$ of linearly independent vectors $\bb{m}_i\in\RR^N$ $(N\ge n)$, and define a subspace $\mathcal{M}$ of $\RR^N$ as, $\mathcal{M}=\spn\{\bb{m}_1,\bb{m}_2,\linebreak\dots,\bb{m}_n\}$. 
For any non-zero cue $\bb{m}_c$, the solution of Eq. (\ref{system_retrieval}) under input (\ref{cue_input}) asymptotically approaches to some periodic solution $\bb{x}_r^*(t)$. Especially if $\bb{m}_c\not\in S_\perp$, $\bb{x}_r^*(t)$ becomes periodically retrievable at $t=t^\dagger>0$ where
\begin{equation}\label{t_dagger}
t^\dagger = \frac{1}{\omega}\tan^{-1}\omega + n\frac{\pi}{\omega},\quad n\in\ZZ.
\end{equation} 
%then the followings are also true:
%\begin{itemize}[itemsep=-2pt]
%\item[\textbf{\textit{a.}}] If $\bb{m}_c$ was retrievable, then $\bb{x}_r^*(t)$ is retrievable for all $t>0$.
%\item[\textbf{\textit{b.}}] If $\bb{m}_c$ was non-retrievable, then $\bb{x}_r^*(t)$ at least becomes retrievable on $t=t^\dagger$ when $\bb{m}_c\not\in S_\perp$.
%\end{itemize}

%\begin{enumerate}[itemsep=-2pt]
%\item If $\bb{m}_c\in S_\perp^c\cap\mathcal{M}^c$, then only for $t=t^\dagger$, $\bb{x}_r^*(t)\in\mathcal{M}$, and in fact $\bb{x}_r^*(t^\dagger)\in S$.
%\item If $\bb{m}_c\in S_{\perp}^c\cap\mathcal{M}$, then  $\forall t>0$, $\bb{x}_r^*(t)\in\mathcal{M}$, and in fact $\bb{x}_r^*(t^\dagger)\in S$.
%\item If $\bb{m}_c\in S_{\perp}\cap\mathcal{M}^c$, then $\forall t>0$, $\bb{x}_r^*(t)\not\in\mathcal{M}$ and in fact $\bb{x}^*_r(t^\dagger)=\bb{0}$.
%\item If $\bb{m}_c\in S_\perp\cap\mathcal{M}$, then $\forall t>0$ except $t = t^\dagger$, $\bb{x}_r^*(t)\in\mathcal{M}$, and in fact $\bb{x}^*_r(t^\dagger)=\bb{0}$.
%\end{enumerate}
\end{thm}
\end{mdframed}
Note that, since the retrieval dynamics $\bb{x}_r(t)$ is attracted to a limit cycle $\bb{x}^*_r(t),$ its retrievability depends on that of  $\bb{x}^*_r(t).$ 
The minimal condition for the retrievablity mentioned in Theorem 2 can be extended further:  the proximity of $\bb{m}_c$ to $S$ and $\mathcal{M}$ determines the retrievability of $\bb{x}^*_r(t)$ as follows.
\begin{enumerate}[label=(\roman*),itemsep=-3pt]
     \item Case $\bb{m}_c\in\mathcal{M}$ (good cue): $\bb{x}^*_r(t)\in\mathcal{M}$ for all $t$.
     \item Case $\bb{m}_c\in S_\perp^c\cap\mathcal{M}^c$ (relavent cue): $\bb{x}^*_r(t)$ is retrievable at $t=t^\dagger$ as in Theorem 2.
     \item Case $\bb{m}_c\in S_\perp\cap\mathcal{M}^c$ (wrong cue): $\bb{x}_r^*(t)$ never becomes retrievable.
\end{enumerate}
 Fig. \ref{fig_illust} gives a graphical illustration about dependence of the retrieval dynamics on the memory cue.  More details about the retrievability conditions incuding the proof of Theorem \ref{retrieval_thm} can be found in Appendix B3.

%Lastly, in terms of associative memory, one's point of interest naturally considers the case of \textit{relevant} memory cues, that is, $\bb{m}_c$ behaving near to the set of original representations $\{\bb{m}_i\}_{i=1}^n$. According to Theorem 2, S should be well designed to ensure that such relevant cues do not become perpendicular to itself and fail the retrieval. 

From the above analysis, one can see that the chance for good and relavent cues increases 
if the memory plane $S$ is formed near the memory representations $\bb{m}_1,\cdots,\bb{m}_n$. 
To measure the distance between $S$ and each memory representation $\bb{m}_i$, one can use the mean cosine similarity
$\langle\cos\theta_i\rangle_i$
where $\theta_i$ represents the angle between each $\bb{m}_i$ and $S$.
Note that if  $\langle\cos\theta_i\rangle_i=1$ if all $\bb{m}_1,\cdots,\bb{m}_n$ are embedded in $S$.
The next theorem tells that one can choose the optimal sampling time for input $\xi_1,\dots,\xi_n$ in Eq. (\ref{memory_input1}).

\begin{mdframed}[
backgroundcolor = grisframe,
outerlinewidth = 0.7pt,
innertopmargin = -4pt, 
leftmargin = 10pt, 
rightmargin = 10pt,
]
\begin{thm}
\name{Optimal Choice for $\xi_i$}
Suppose $\{\bb{m}_i\}_{i=1}^n$ with $\bb{m}_i\in\RR^N$ are mutually orthogonal vectors of the same magnitude.
Then the maximum value of $\langle\cos\theta_i\rangle_i$ is  $\sqrt{\frac{2}{n}}$ and can be attained with the
the distribution of $\xi_i$ as
\begin{equation}\label{max_perf}
\xi_i=(i-1)\frac{\pi}{n} + \alpha,\quad i=1,\dots,n, \;0\le\alpha<\frac{\pi}{n}.
\end{equation}
\end{thm}
\end{mdframed}
The theorem suggests a uniform sampling times $\xi_i$ for the sequential input of representations in order to naturally maximize the expected performance of retrieval. 
%As a remark, for above such choices (\ref{xi_eq}) of $\xi_i$, one can analogously deduce that for generally given set $\{\bb{m}_i\}_{i=1}^n$ (not necessarily mutually orthogonal), each $\theta_i(\xi_1,\dots,\xi_n)$ between resulting $S(\xi_1,\dots,\xi_n)$ and respective $\bb{m}_i$ must satisfy
%\begin{equation}
%\abs{\theta_i(\xi_1,\dots,\xi_n)} \le \cos^{-1}\sbr{\sqrt{\frac{2}{n}}},\quad i=1,\dots,n.
%\end{equation}
%
%Table 1 shows the value of $\max\limits_{\xi_1,\dots,\xi_n} \langle\cos\abs{\theta_i(\xi_1,\dots,\xi_n)}\rangle_i$ for some $n\ge 2$ in the case of orthonormal set $\{\bb{m}_i\}_{i=1}^n$.
%
%\begin{table}[h!]\small \renewcommand{\arraystretch}{1.2}
%\centering
%\setlength{\tabcolsep}{4.7pt}
%\begin{tabular}{ | c | c | c | c | c | c | c | c | c | c | }
%\hline
%$n\;(\ge 2)$ & 2 & 3 & 4 & 5 & 6 & 7 & 8 & 9 & 10 \\
%\hline
%$\max_{\xi_1,\dots,\xi_n}\langle\cos\abs{\theta_i}\rangle_i$ & 1.0000 & 0.8165 & 0.7071  &  0.6325  &  0.5774 & 0.5345  &  0.5000  &  0.4714  &  0.4472 \\
%\hline
%\end{tabular}\caption{\small  The value of $\max_{\xi_1,\dots,\xi_n} \langle\cos\abs{\theta_i(\xi_1,\dots,\xi_n)}\rangle_i$ for some $n\;(\ge 2)$, in the case of orthonormal set $\{\bb{m}_i\}_{i=1}^n$. The higher the value, the better the expected performance in memory retrieval of the memory plane $S(\xi_1,\dots,\xi_n)$, respect to each original memory representations $\bb{m}_i$.  }
%\end{table}
%
\section*{Discussion}
There are now substantial evidences accumulated that such macroscopic neural oscillations are related to memory encoding, attention, and integration of visual patterns \cite{singer1995visual, gupta2016oscillatory, rutishauser2010human}. 
Our analysis supports such functional role of neural oscillations, by veiwing tham as limit cyles related to a memory plane which stores the information in the form of  an anti-symmetric connectivity.
We were able to show that the performance of retrieval is determined by the similarity of a memory cue to the original data. This suggests an alternative computational approach that can handle high dimensional and strongly associated data sets from a biomimetic perspective.  

The cognitive systems do not simply receive an external input in a passive way, 
but rather actively pose it on acceptance.
It is therefore reasonable to assume that there is some pre-encoding process to encode the external inputs, say, $\bb{f}_1,\dots,\bb{f}_n$ into the memory components $\bb{m}_1,\dots,\bb{m}_n$ in the neural state space. To model such preprocess,
one can use a set of internal tag vectors $\bb{r}_1,\dots,\bb{r}_n$. 
For example, one of possible ways of encoding is to use tensor product like $\bb{m}_i=\bb{f}_i\otimes\bb{r}_i.$
Then the tag vectors can be also used for decoding $\bb{x}(t)$, that is, to retreive the external inputs $\bb{f}_1,\dots,\bb{f}_n$ from $\bb{x}(t)$, while $\bb{x}(t)$ is retrievable.
In the separate paper, we will study  the end-to-end memory process with encoding/decoding processes, focusing on more practical issues such as how the network can embed actual data into neural representations for efficient reproduction from the retrievable states.
 
\subsection*{Acknowledgements}
P. Kim was supported by National Research Foundation of Korea (2017R1D1A1B04032921) and H. Yoon was supported by Ulsan National Institute of Science and Technology \linebreak 12(1.200052.01).

%\small
\subsection*{Appendix A: Derivation of the STDP Learning Rule}

Actually, Eq. (\ref{STDP}) in the main text can be equivalently written by the following expression only using convolution defined on $\RR$, i.e.,
\begin{align}\label{STDP2}
\dot{W}_{ij}(t) = -\gamma W_{ij}(t) + \rho\sbr{(K_{\Delta t>0}*\phi(x_j))(t)\phi(x_i)(t) + (K_{\Delta t\le 0}*\phi(x_i))(t)\phi(x_j)(t) },
\end{align}
where $K_{\Delta t>0}(s)$ comes from the kernel $K(s)$ with $s>0$ only, and $K_{\Delta t\le 0}(s)$ is $K(-s)$ with $s\le 0$ only. Therefore, in the case of kernels only behaving as Eq. (\ref{kernel}), we have $K_{\Delta t>0}(s) = \delta(s-s_0)=\delta_{s_0}(s)$ and $K_{\Delta t\le 0}(s) = -\delta(-s+s_0) = -\delta(s-s_0)=-\delta_{s_0}(s)$.
Now, since $(\delta_{s_0}*f)(t)=\int_\RR \delta(s-s_0)f(t-s)\;\dd s = f(t-s_0)$, thus the terms in Eq. (\ref{STDP2}) including convolution is simplified into
\begin{equation}
\dot{W}_{ij}(t) = -\gamma W_{ij}(t) + \rho\sbr{\phi(x_j(t-s_0))\phi(x_i(t)) - \phi(x_i(t-s_0))\phi(x_j(t))}.
\end{equation}
Rewriting it in matrix form,
\begin{equation}\label{eq_diracconv}
\dot{\bb{W}}(t) = -\gamma \bb{W}(t) + \rho\sbr{\phi(\bb{x}(t))\phi(\bb{x}(t-s_0))^\top-\phi(\bb{x}(t-s_0))\phi(\bb{x}(t))^\top}.
\end{equation}
Now, if one specifies $s_0$ with some $s_0=\tau>0$ and applies approximation $\phi(x)\approx x$, then Eq. (\ref{eq_diracconv}) concisely reduces to
\begin{equation}\label{system_full}
\dot{\bb{W}} = -\gamma \bb{W} + \rho(\bb{x}\bb{x}_\tau^\top-\bb{x}_\tau\bb{x}^\top)
\end{equation}
with introducing notation $\bb{x}_\tau=\bb{x}(t-\tau)$ as in the main text. This is our evolution equation on $\dot{\bb{W}}$, where the term $-\gamma\bb{W}$ acts as homeostatic decay and $\rho(\bb{x}\bb{x}_\tau^\top-\bb{x}_\tau\bb{x}^\top)$ acts as actual learning operator by STDP. \hfill$\blacksquare$

\subsection*{Appendix B: Proofs for the Theoretical Results}

\subsubsection*{\textit{B1: Proof of Lemma A}}
Let $\bb{b}(t)=[b_1(t)\;\cdots\;b_N(t)]^\top$, and $\bb{m}_i = [m_{i_1}\;\dots\;m_{i_N}]^\top$. Then, each component of $\bb{b}(t)$ satisfies 
\begin{align}
b_j(t) & = \sum_{i=1}^{n}m_{i_j}\sin(\omega t-\xi_i) \notag \\
& = \sum_{i=1}^{n}m_{i_j}(\sin\omega t\cos\xi_i-\cos\omega t\sin\xi_i) \notag\\
& = \cos\omega t\sbr{\sum_{i}^{n}m_{i_j}(-\sin\xi_i)} + \sin\omega t\sbr{\sum_{i}^{n}m_{i_j}\cos\xi_i},\;\quad j=1,\dots,N. \label{sinbreak}
\end{align}
Thus if we introduce
\begin{equation}
\begin{dcases}
\bb{u} = -\begin{bmatrix}
\sum_{i=1}^n m_{i_1}\sin\xi_i&\cdots & \sum_{i=1}^n m_{i_N}\sin\xi_i
\end{bmatrix}^\top \\
\bb{v} = \begin{bmatrix}
\sum_{i=1}^n m_{i_1}\cos\xi_i & \cdots & \sum_{i=1}^n m_{i_N}\cos\xi_i
\end{bmatrix}
^\top,
\end{dcases}
\end{equation} 
then this choice of $\bb{u},\bb{v}$ can be represented in alternate form of $\bb{u}=-\boldsymbol\Psi\sin\boldsymbol\xi$ and $\bb{v} = \boldsymbol\Psi\cos\boldsymbol\xi$ where $\boldsymbol\Psi$, $\sin\boldsymbol\xi$, and $\cos\boldsymbol\xi$ are defined as in the theorem statement, and guarantees 
\begin{equation}\label{bt_plane}
\bb{b}(t) = \cos\omega t\bb{u}+\sin\omega t\bb{v}
\end{equation} 
by Eq. (\ref{sinbreak}). Therefore $\bb{b}(t)$ is periodic and embedded in plane $\spn\{\bb{u},\bb{v}\}$. \hfill$\blacksquare$

\subsubsection*{\textit{B2: Proof of Theorem 1}}
The proof of this theorem requires the following lemma which describes some algebraic relations of frequently appearing periodic functions in the behavior of neural periodic solution $\bb{x}^*(t)$.

\begin{mdframed}[
backgroundcolor = grisframe,
outerlinewidth = 0.7pt,
innertopmargin = -4pt, 
leftmargin = 10pt, 
rightmargin = 10pt,
]
\begin{customlem}{B}\label{ci_lemma}
Let $c_{i,[\lambda,\omega]}(t)$ be the following periodic functions in $t$, with parameters $\lambda,\omega\in\RR^{+}$, which is defined as
\begin{equation}\label{thm_limitingcoeff2}
\begin{dcases}
c_{1,[\lambda,\omega]}(t) = \frac{1}{2}\sbr{\frac{\cos(\omega t-\theta_{-,[\lambda,\omega]})}{\sqrt{\Phi_{-,[\lambda,\omega]}}}+\frac{\cos(\omega t-\theta_{+,[\lambda,\omega]})}{\sqrt{\Phi_{+,[\lambda,\omega]}}}}  \\
c_{2,[\lambda,\omega]}(t) = \frac{1}{2}\sbr{\frac{\sin(\omega t-\theta_{-,[\lambda,\omega]})}{\sqrt{\Phi_{-,[\lambda,\omega]}}}-\frac{\sin(\omega t-\theta_{+,[\lambda,\omega]})}{\sqrt{\Phi_{+,[\lambda,\omega]}}}}  \\
c_{3,[\lambda,\omega]}(t) = \frac{1}{2}\sbr{\frac{\sin(\omega t-\theta_{-,[\lambda,\omega]})}{\sqrt{\Phi_{-,[\lambda,\omega]}}}+\frac{\sin(\omega t-\theta_{+,[\lambda,\omega]})}{\sqrt{\Phi_{+,[\lambda,\omega]}}}} \\
c_{4,[\lambda,\omega]}(t) = -\frac{1}{2}\sbr{\frac{\cos(\omega t-\theta_{-,[\lambda,\omega]})}{\sqrt{\Phi_{-,[\lambda,\omega]}}}+\frac{\cos(\omega t-\theta_{+,[\lambda,\omega]})}{\sqrt{\Phi_{+,[\lambda,\omega]}}}},
\end{dcases}
\end{equation}
where $\theta_{\pm,[\lambda,\omega]} = \tan^{-1}(\omega\pm\lambda)$, $\Phi_{\pm,[\lambda,\omega]}=\lambda^2\pm 2\omega\lambda + \omega^2+ 1$. Let's denote $\frac{\dd }{\dd t}c_{i,[\lambda,\omega]}(t)$ with $\dot{c}_{i,[\lambda,\omega]}(t)$. 
Then, the followings are true:
\begin{itemize}[itemsep=0pt]
\item[\textbf{1.}] $\dot{c}_{1,[\lambda,\omega]} = -\omega c_{3,[\lambda,\omega]}$, $\dot{c}_{2,[\lambda,\omega]}=-\omega c_{4,[\lambda,\omega]}$, $\dot{c}_{3,[\lambda,\omega]}=\omega c_{1,[\lambda,\omega]}$, and $\dot{c}_{4,[\lambda,\omega]}=\omega c_{2,[\lambda,\omega]}$. 
\item[\textbf{2.}] \begin{itemize}[itemsep=0pt]
\item[\textbf{a.}] $c_{1,[\lambda,\omega]}(t)-\omega c_{3,[\lambda,\omega]}(t)+\lambda c_{2,[\lambda,\omega]}(t) = \cos\omega t$.
\item[\textbf{b.}] $c_{1,[\lambda,\omega]}(t)+\omega c_{3,[\lambda,\omega]}(t)+\lambda c_{4,[\lambda,\omega]}(t) = \sin\omega t$.
\item[\textbf{c.}] $c_{2,[\lambda,\omega]}(t)-\omega c_{4,[\lambda,\omega]}(t)-\lambda c_{1,[\lambda,\omega]}(t) = 0$.
\item[\textbf{d.}] $c_{2,[\lambda,\omega]}(t)+\omega c_{4,[\lambda,\omega]}(t)-\lambda c_{3,[\lambda,\omega]}(t) = 0$.
\end{itemize}
\end{itemize}
\end{customlem}
\end{mdframed}
\begin{proof}
Statement \textit{1} can be straightforwardly shown by direct differentiation. For statement \textit{2}, omitting the $_{[\lambda,\omega]}$ notations in $\theta_{\pm,[\lambda,\omega]}$, observe that 
\begin{align}
& c_{1,[\lambda,\omega]}(t)-\omega c_{3,[\lambda,\omega]}(t)+\lambda c_{2,[\lambda,\omega]}(t) \notag \\
& =  \frac{1}{2}\!\left(\frac{\cos(\omega t-\theta_-)-(\omega-\lambda)\sin(\omega t-\theta_-)}{\sqrt{\Phi_{-,[\lambda,\omega]}}}+\frac{\cos(\omega t-\theta_+)-(\omega+\lambda)\sin(\omega t-\theta_+)}{\sqrt{\Phi_{+,[\lambda,\omega]}}}\right) \label{lemmaB1}\\
& = \frac{1}{2}\sbr{\frac{\sqrt{\Phi_{-,[\lambda,\omega]}}\cos(\omega t-\theta_-+\theta_-)}{\sqrt{\Phi_{-,[\lambda,\omega]}}}+\frac{\sqrt{\Phi_{+,[\lambda,\omega]}}\cos(\omega t-\theta_++\theta_+)}{\sqrt{\Phi_{+,[\lambda,\omega]}}}} \label{lemmaB2}\\
& = \cos\omega t,\notag
\end{align}
where Eq. (\ref{lemmaB1}) comes from direct substitution and Eq. (\ref{lemmaB2}) comes from the fact $1^2+(\omega\pm\lambda)^2=\Phi_{\pm,[\lambda,\omega]}$ and $a\cos t-b\sin t=\sqrt{a^2+b^2}\cos(t+\phi)$ with $\phi=-\tan^{-1}(a/b)$, so \textit{2a} has been shown. Differentiating both sides of this result respect to $t$ using statement \textit{1} directly yields $c_{1,[\lambda,\omega]}(t)+\omega c_{3,[\lambda,\omega]}(t)+\lambda c_{4,[\lambda,\omega]}(t) = \sin\omega t$, which is \textit{2b}. Similarly, one can also check \textit{2c} by
\begin{align}
& c_{2,[\lambda,\omega]}(t)-\omega c_{4,[\lambda,\omega]}(t)-\lambda c_{1,[\lambda,\omega]}(t) \notag \\
& = \frac{1}{2}\sbr{\frac{\sin(\omega t-\theta_-)+(\omega-\lambda)\cos(\omega t-\theta_-)}{\sqrt{\Phi_{-,[\lambda,\omega]}}}-\frac{\sin(\omega t-\theta_+)+(\omega+\lambda)\cos(\omega t-\theta_+)}{\sqrt{\Phi_{+,[\lambda,\omega]}}}} \notag\\
& = \frac{1}{2}\sbr{\frac{\sqrt{\Phi_{-,[\lambda,\omega]}}\sin(\omega t-\theta_-+\theta_-)}{\sqrt{\Phi_{-,[\lambda,\omega]}}}-\frac{\sqrt{\Phi_{+,[\lambda,\omega]}}\sin(\omega t-\theta_++\theta_+)}{\sqrt{\Phi_{+,[\lambda,\omega]}}}} \notag\\
& = 0,\label{ci_0}
\end{align}
and differentiating both sides of this result respect to $t$ using statement \textit{1} also yields $c_{2,[\lambda,\omega]}(t)+\omega c_{4,[\lambda,\omega]}(t)-\lambda c_{3,[\lambda,\omega]}(t) = 0$, which is statement \textit{2d}. \nopagebreak
\end{proof}

Now, we can proceed to the proof of Theorem \ref{thm_1}.\medskip

\noindent\textit{\textbf{Proof}}.
Let $\bb{x}^*(t)$ and $\bb{W}^*$ be the corresponding periodic solution with steady synapse in Theorem \ref{thm_1}. Let $c_{i,[\lambda,\omega]}(t)$ and $\Phi_{\pm,[\lambda,\omega]}$ be the periodic functions  with parameters $\lambda$, $\omega$ and polynomials in $\lambda$ defined as in Lemma \ref{ci_lemma} respectively. Point-blank, we propose the followings: 
\begin{mdframed}[
backgroundcolor = grisframe,
outerlinewidth = 0.7pt,
leftmargin = 10pt, 
rightmargin = 10pt,
]
The solution pair $(\bb{x}^*(t),\bb{W}^*)$ are given by
\begin{equation}\label{thm_sol}
\begin{dcases}
\bb{x}^*(t) = f(t)\bb{u} + g(t)\bb{v} \\
\bb{W}^* = \alpha(\bb{v}\bb{u}^\top-\bb{u}\bb{v}^\top)
\end{dcases}
\end{equation}
where $\alpha\in\RR$, and $f,g:\RR\to\RR$ periodic functions, and vectors $\bb{u}$, $\bb{v}$ given as a form in Lemma \ref{lem_plane} (i.e., $\bb{u}=-\boldsymbol\Psi\sin\boldsymbol\xi$, $\bb{v}=\boldsymbol\Psi\cos\boldsymbol\xi$). Especially, $\alpha$, $f(t)$ and $g(t)$ are given by 
\begin{equation}\label{thm_sol_values}
\begin{dcases}
f(t) = c_{1,[\lambda_0,\omega]}(t)-\frac{\mu}{\sqrt{1-\mu^2}}c_{2,[\lambda_0,\omega]}(t)-\frac{\eta_2}{\eta_1\sqrt{1-\mu^2}}c_{4,[\lambda_0,\omega]}(t) \\
g(t) = c_{3,[\lambda_0,\omega]}(t)+\frac{\mu}{\sqrt{1-\mu^2}}c_{4,[\lambda_0,\omega]}(t) +\frac{\eta_1}{\eta_2\sqrt{1-\mu^2}}c_{2,[\lambda_0,\omega]}(t) \\
\alpha = \frac{\lambda_0}{{\eta_1\eta_2}\sqrt{1-\mu^2}},
\end{dcases}
\end{equation}
where $\lambda_0$ is a real root of algebraic equation $h(\lambda)=0$ with
\begin{equation}\label{thm_lambdacond}
h(\lambda) = \frac{\lambda\,\Phi_{-,[\lambda,\omega]}\Phi_{+,[\lambda,\omega]}}{\sbr{\eta_1\eta_2\sqrt{1-\mu^2}}(\lambda^2+\omega^2+1)+(\eta_1^2+\eta_2^2)\omega\lambda}-\frac{\rho\sin\omega\tau}{\gamma},
\end{equation}
and constants $\eta_1$, $\eta_2$, and $\mu$ are
\begin{equation}\label{thm_etacond}
\eta_1=\norm{\bb{u}},\;\eta_2=\norm{\bb{v}},\;\text{and}\;\mu = \frac{\bb{u}^\top\bb{v}}{\norm{\bb{u}}\norm{\bb{v}}}.
\end{equation}
\end{mdframed}

To begin with, we will show that the solution pair $(\bb{x}^*(t),\bb{W}^*)$ in Eq. (\ref{thm_sol}) with condition (\ref{thm_sol_values}), (\ref{thm_lambdacond}), and (\ref{thm_etacond}) satisfies Eq. (\ref{system}). First, let's start with showing $\dot{\bb{x}}^*+\bb{x}^*-\bb{W}^*\bb{x}^* = \bb{b}(t)$. According to Lemma \ref{lem_plane}, such $\bb{b}(t)$ of form (\ref{memory_input1}) is equivalent with $\cos\omega t\bb{u}+\sin\omega t\bb{v}$ on plane (see Eq. (\ref{bt_plane})) $S$, so it only requires checking $\dot{\bb{x}}^*+\bb{x}^*-\bb{W}^*\bb{x}^* = \cos\omega t\bb{u}+\sin\omega t\bb{v}$. 

For this, from Lemma \ref{ci_lemma}-\textit{1}, firstly see that $\dot{c}_{1,[\lambda_0,\omega]} = -\omega c_{3,[\lambda_0,\omega]}$, $\dot{c}_{2,[\lambda_0,\omega]}=-\omega c_{4,[\lambda_0,\omega]}$, $\dot{c}_{3,[\lambda_0,\omega]}=\omega c_{1,[\lambda_0,\omega]}$, and $\dot{c}_{4,[\lambda_0,\omega]}=\omega c_{2,[\lambda_0,\omega]}$. Thus from complete expression of $\bb{x}^*(t)$,
\begin{equation}
\begin{split}
\dot{\bb{x}}^*(t)+\bb{x}^*(t) & = \sbr{(c_1-\omega c_3) + \frac{\mu}{\sqrt{1-\mu^2}}(-c_2+\omega c_4) +\frac{\eta_2}{\eta_1\sqrt{1-\mu^2}}(-c_2-\omega c_4) }\bb{u} \\
&\quad\quad + \sbr{(c_1+\omega c_3) + \frac{\mu}{\sqrt{1-\mu^2}}(c_2+\omega c_4) +\frac{\eta_1}{\eta_2\sqrt{1-\mu^2}}(c_2-\omega c_4) }\bb{v},
\end{split}
\end{equation}
where $_{[\lambda_0,\omega]}$ notations in $c_{i,[\lambda_0,\omega]}$ are omitted. 

On the other hand, for remaining computations, we introduce some additional definitions in order to make the following processes concise. Set $\bb{u}_\perp$ and $\bb{v}_\perp$ as
\begin{equation}\label{thm_perpexpression}
\bb{u}_\perp = \frac{1}{\sqrt{1-\mu^2}}\sbr{-\mu\bb{u}+\frac{\eta_1}{\eta_2}\bb{v}},\quad\bb{v}_\perp = \frac{1}{\sqrt{1-\mu^2}}\sbr{-\frac{\eta_2}{\eta_1}\bb{u}+\mu\bb{v}},
\end{equation}
where $\eta_1=\norm{\bb{u}}$, $\eta_2=\norm{\bb{v}}$ and $\mu = \bb{u}^\top\bb{v}/\sbr{\norm{\bb{u}}\norm{\bb{v}}}$. Then one can see that $\bb{u}_\perp\perp\bb{u}$ satisfying $\norm{\bb{u}_\perp}=\norm{\bb{u}}$ and $\bb{v}_\perp\perp\bb{v}$ satisfying $\norm{\bb{v}_\perp}=\norm{\bb{v}}$, and can further check that the expression for $\bb{x}^*(t)$ is equivalent with $c_{1,[\lambda_0,\omega]}(t)\bb{u}+c_{2,[\lambda_0,\omega]}(t)\bb{u}_\perp+c_{3,[\lambda_0,\omega]}(t)\bb{v}+c_{4,[\lambda_0,\omega]}(t)\bb{v}_\perp$. Now performing computation of $\bb{W}^*\bb{x}^*(t)$ yields
\begingroup
\allowdisplaybreaks
\begin{align}
\bb{W}^*\bb{x}^*(t) 
%& = \frac{\lambda_0}{\eta_1\eta_2\sqrt{1-\mu^2}}(\twist{\bb{v}}{\bb{u}})\sbr{ c_{1,[\lambda_0,\omega]}\bb{u}+c_{2,[\lambda_0,\omega]}\bb{u}_\perp+c_{3,[\lambda_0,\omega]}\bb{v}+c_{4,[\lambda_0,\omega]}\bb{v}_\perp } \notag \\
%& = \frac{\lambda_0}{\eta_1\eta_2\sqrt{1-\mu^2}}\left( (\eta_1^2 c_{1,[\lambda_0,\omega]}+\eta_1\eta_2\mu c_{3,[\lambda_0,\omega]}-\eta_1\eta_2\sqrt{1-\mu^2}c_{4,[\lambda_0,\omega]})\bb{v} \right. \notag \\
%& \quad\quad\quad\quad\quad\quad\quad +\left.(-\eta_1\eta_2\mu c_{1,[\lambda_0,\omega]}-\eta_1\eta_2\sqrt{1-\mu^2}c_{2,[\lambda_0,\omega]}-\eta_2 c_{3,[\lambda_0,\omega]})\bb{u} \right)\\
& = -\lambda_0\sbr{c_{2,[\lambda_0,\omega]}+\frac{\mu}{\sqrt{1-\mu^2}}c_{1,[\lambda_0,\omega]}+\frac{\eta_2}{\eta_1\sqrt{1-\mu^2}}c_{3,[\lambda_0,\omega]} }\bb{u} \notag\\
& \quad\quad\quad\quad +\lambda_0\sbr{-c_{4,[\lambda_0,\omega]} +\frac{\mu}{\sqrt{1-\mu^2}}c_{3,[\lambda_0,\omega]} + \frac{\eta_1}{\eta_2\sqrt{1-\mu^2}}c_{1,[\lambda_0,\omega]} }\bb{v}, \notag
\end{align}
\endgroup
where we used the facts $\bb{u}^\top\bb{u}=\eta_1^2$, $\bb{v}^\top\bb{v}=\eta_2^2$, $\bb{u}^\top\bb{v} = \eta_1\eta_2\mu$, $\bb{u}^\top\bb{v}_\perp = -\eta_1\eta_2\sqrt{1-\mu^2}$, and $\bb{v}^\top\bb{u}_\perp=+\eta_1\eta_2\sqrt{1-\mu^2}$.
Now combining above results with full notations, we have
\begingroup
\allowdisplaybreaks
\begin{align}
& \dot{\bb{x}}^*(t)+\bb{x}^*(t)-\bb{W}^*\bb{x}^*(t)\notag \\
& = \left( (c_{1,[\lambda_0,\omega]}-\omega c_{3,[\lambda_0,\omega]}+\lambda_0c_{2,[\lambda_0,\omega]}) - (c_{2,[\lambda_0,\omega]}-\omega c_{4,[\lambda_0,\omega]}-\lambda_0c_{1,[\lambda_0,\omega]})\frac{\mu}{\sqrt{1-\mu^2}}\right.\notag \\
& \quad\quad\quad\quad\quad\quad\quad\quad\quad\quad\quad\quad \left.-(c_{2,[\lambda_0,\omega]}+\omega c_{4,[\lambda_0,\omega]}-\lambda_0c_{3,[\lambda_0,\omega]})\frac{\eta_2}{\eta_1\sqrt{1-\mu^2}} \right)\bb{u} \label{22}\\
& \quad+ \left( (c_{1,[\lambda_0,\omega]}+\omega c_{3,[\lambda_0,\omega]}+\lambda_0c_{4,[\lambda_0,\omega]}) +(c_{2,[\lambda_0,\omega]}+\omega c_{4,[\lambda_0,\omega]}-\lambda_0c_{3,[\lambda_0,\omega]})\frac{\mu}{\sqrt{1-\mu^2}} \right.\notag \\
& \quad\quad\quad\quad\quad\quad\quad\quad\quad\quad\quad\quad \left.+(c_{2,[\lambda_0,\omega]}-\omega c_{4,[\lambda_0,\omega]}-\lambda_0c_{1,[\lambda_0,\omega]})\frac{\eta_1}{\eta_2\sqrt{1-\mu^2}} \right)\bb{v}. \label{23}
\end{align}
\endgroup
Now, Lemma \ref{ci_lemma}-\textit{2a}, \textit{2c}, and \textit{2d} tells us that each three coefficients in terms of $c_{i,[\lambda_0,\omega]}(t)$ of (\ref{22}) in RHS is $\cos\omega t$, $0$, and $0$ respectively, thus simplified only into $\cos\omega t \bb{u}$. Similarly, each three coefficients in terms of $c_{i,[\lambda_0,\omega]}(t)$ in (\ref{23}) becomes $\sin\omega t$, $0$, and $0$ by Lemma \ref{ci_lemma}-\textit{2b}, \textit{2c}, and \textit{2d} respectively, thus yielding $\sin\omega t\bb{v}$. Therefore in total, completing the proof of $\dot{\bb{x}}^*(t)+\bb{x}^*(t)-\bb{W}^*\bb{x}^*(t) = \cos \omega t\bb{u}+\sin\omega t\bb{v}=\bb{b}(t)$.

Now, it remains to confirm $-\gamma \bb{W}^*+\rho(\twist{\bb{x}^*}{\bb{x}^*_\tau})= \bb{O}$. In order to show this, first we have to compute the term $\twist{\bb{x}^*}{\bb{x}^*_\tau}$. Putting $\bb{x}(t) =c_{1,[\lambda_0,\omega]}(t)\bb{u}+c_{2,[\lambda_0,\omega]}(t)\bb{u}_\perp+c_{3,[\lambda_0,\omega]}(t)\bb{v}+c_{4,[\lambda_0,\omega]}(t)\bb{v}_\perp$ and doing some lengthy computations with the help of following trigonometric relations
\begin{equation}
\begin{dcases}
\cos(\omega t-\theta_-)\sin(\omega t-\theta_+)_\tau-\cos(\omega t-\theta_-)_\tau\sin(\omega t-\theta_+) = -\sin\omega\tau\cos(\theta_--\theta_+) \\
\sin(\omega t-\theta_-)\sin(\omega t-\theta_+)_\tau-\sin(\omega t-\theta_-)_\tau\sin(\omega t-\theta_+) = \sin\omega\tau\sin(\theta_--\theta_+) \\
\sin(\omega t-\theta_-)\cos(\omega t-\theta_+)_\tau-\sin(\omega t-\theta_-)_\tau\cos(\omega t-\theta_+) = \sin\omega\tau\cos(\theta_--\theta_+) \\
\cos(\omega t-\theta_-)\cos(\omega t-\theta_+)_\tau-\cos(\omega t-\theta_-)_\tau\cos(\omega t-\theta_+) = \sin\omega\tau\sin(\theta_--\theta_+),
\end{dcases}
\end{equation} 
then one gets
\begin{align}
& \bb{x}^*(t)\bb{x}^*(t-\tau)^\top-\bb{x}^*(t-\tau)\bb{x}^*(t)^\top \notag \\
& \quad= \frac{\sin\omega\tau}{4}\left[\sbr{\frac{1}{\Phi_{-,[\lambda_0,\omega]}}-\frac{1}{\Phi_{+,[\lambda_0,\omega]}}}\sbr{\sbr{\bb{u}_\perp\bb{u}^\top-\bb{u}\bb{u}_\perp^\top}+\sbr{\bb{v}_\perp\bb{v}^\top-\bb{v}\bb{v}_\perp^\top}}\right.\notag \\ 
& \quad\quad\quad+\sbr{\frac{1}{\Phi_{-,[\lambda_0,\omega]}}+\frac{1}{\Phi_{+,[\lambda_0,\omega]}} +\frac{2\cos(\theta_{-,[\lambda_0,\omega]} -\theta_{+,[\lambda_0,\omega]})}{\Phi_{-,[\lambda_0,\omega]}\Phi_{+,[\lambda_0,\omega]}}}\sbr{\bb{v}\bb{u}^\top-\bb{u}\bb{v}^\top}\notag  \\
& \quad\quad\quad+\sbr{\frac{1}{\Phi_{-,[\lambda_0,\omega]}}+\frac{1}{\Phi_{+,[\lambda_0,\omega]}} -\frac{2\cos(\theta_{-,[\lambda_0,\omega]} -\theta_{+,[\lambda_0,\omega]})}{\Phi_{-,[\lambda_0,\omega]}\Phi_{+,[\lambda_0,\omega]}}}\sbr{\bb{v}_\perp\bb{u}_\perp^\top-\bb{u}_\perp\bb{v}_\perp^\top}\notag  \\
& \quad\quad\quad\left.+\frac{2\sin(\theta_{-,[\lambda_0,\omega]} -\theta_{+,[\lambda_0,\omega]})}{\Phi_{-,[\lambda_0,\omega]}\Phi_{+,[\lambda_0,\omega]}}\sbr{\sbr{\bb{v}_\perp\bb{u}^\top-\bb{u}\bb{v}_\perp^\top}+\sbr{\bb{v}\bb{u}_\perp^\top-\bb{u}_\perp\bb{v}^\top}}\right], \label{thm_twistcomp}
\end{align}
which is a constant in $t$ as expected. Here, from (\ref{thm_perpexpression}), one can easily check that the six anti-symmetric matrix terms in (\ref{thm_twistcomp}) satisfy the following relations: 
\begin{equation}\label{thm_relation}
\begin{dcases}
\bb{u}_\perp\bb{u}^\top-\bb{u}\bb{u}_\perp^\top = \frac{\eta_1}{\eta_2\sqrt{1-\mu^2}}\sbr{\bb{v}\bb{u}^\top-\bb{u}\bb{v}^\top} \\
\bb{v}_\perp\bb{v}^\top-\bb{v}\bb{v}_\perp^\top = \frac{\eta_2}{\eta_1\sqrt{1-\mu^2}}\sbr{\bb{v}\bb{u}^\top-\bb{u}\bb{v}^\top} \\
\bb{v}_\perp\bb{u}_\perp^\top-\bb{u}_\perp\bb{v}_\perp^\top = \bb{v}\bb{u}^\top-\bb{u}\bb{v}^\top \\
\bb{v}\bb{u}_\perp^\top-\bb{u}_\perp\bb{v}^\top = - \sbr{\bb{v}_\perp\bb{u}^\top-\bb{u}\bb{v}_\perp^\top}.
\end{dcases}
\end{equation}
Now simplifying (\ref{thm_twistcomp}) using (\ref{thm_relation}) in terms of $\twist{\bb{v}}{\bb{u}}$ and substituting the result into $-\gamma \bb{W}^*+\rho(\twist{\bb{x}^*}{\bb{x}^*_\tau})$ alongside substituting $\bb{W}^*=\lambda_0(\twist{\bb{v}}{\bb{u}})/\sbr{\eta_1\eta_2\sqrt{1-\mu^2}}$ together, then
\begin{align}\label{thm_finaleq}
& -\gamma \bb{W}^*+\rho(\twist{\bb{x}^*}{\bb{x}^*_\tau}) \\
& = \left[-\frac{\gamma\lambda_0}{\eta_1\eta_2\sqrt{1-\mu^2}}+\frac{\rho\sin\omega\tau}{\Phi_{-,[\lambda_0,\omega]}\Phi_{+,[\lambda_0,\omega]}}\sbr{\lambda_0^2+\frac{(\eta_1^2+\eta_2^2)\omega}{\eta_1\eta_2\sqrt{1-\mu^2}}\lambda_0+\omega^2+1}\right]\sbr{\bb{v}\bb{u}^\top-\bb{u}\bb{v}^\top}. \notag
\end{align}
Moreover, from the fact that $\lambda_0$ is a root of (\ref{thm_lambdacond}), we know 
\begin{equation}
\frac{\lambda_0\,\Phi_{-,[\lambda_0,\omega]}\Phi_{+,[\lambda_0,\omega]}}{\sbr{\eta_1\eta_2\sqrt{1-\mu^2}}(\lambda_0^2+\omega^2+1)+(\eta_1^2+\eta_2^2)\omega\lambda_0}-\frac{\rho\sin\omega\tau}{\gamma} = 0,
\end{equation} 
and slight more algebra using this shows that the large-bracketed term in (\ref{thm_finaleq}) turns out to be 0, so proving $-\gamma \bb{W}^*+\rho(\twist{\bb{x}^*}{\bb{x}^*_\tau})=\bb{O}$.

From all above, we conclude that Eq. (\ref{thm_sol}) with conditions (\ref{thm_sol_values}), (\ref{thm_lambdacond}), and (\ref{thm_etacond}) is a solution of Eq. (\ref{system}). \hfill$\blacksquare$

\subsubsection*{\textit{B3: Proof of Theorem 2 and the Retrievability Conditions}}

Since Eq. (\ref{system_retrieval}) is a perturbed linear ordinary differential equation, we can obtain explicit solution of $\bb{x}_r(t)$ using the variational formula, i.e.,
\begin{equation}\label{retrieval_soln}
\begin{split}
\bb{x}_r(t) & = e^{t(\bb{W}^*-\bb{I})}\bb{x}_r(0) + \int_{0}^{t}e^{s(\bb{W}^*-\bb{I})}\sin(\omega(t-s))\bb{m}_c\;\dd s \\
& = e^{-t}e^{t\bb{W}^*}\bb{x}_r(0)+\int_{0}^{t}e^{-s}\sin(\omega(t-s))e^{s\bb{W}^*}\bb{m}_c\;\dd s.
\end{split}
\end{equation}

Here, from the fact that $e^{t\bb{W}^*}\bb{x}_0$ is a flow generated by $\dot{\bb{x}}=\bb{W}^*\bb{x}$, $\bb{x}(0)=\bb{x}_0$,  any non-trivial $\bb{x}_0\in S=\spn\{\bb{u},\bb{v}\}$ will generate purely rotational flow on $S$ since $\bb{W}^*=\alpha(\twist{\bb{v}}{\bb{u}})\in\wedge^2(S)$ is rank-2 anti-symmetric. More specifically, there is some $\bb{x}_{0\wedge}\in S$ perpendicular to $\bb{x}_0$ with $\norm{\bb{x}_{0\wedge}}=\norm{\bb{x}_0}$, so that
\begin{equation}\label{antisymm_flow}
e^{t\bb{W}^*}\bb{x}_0=\cos\lambda^* t\bb{x}_0+\sin\lambda^*t\bb{x}_{0\wedge}
\end{equation} 
where $\lambda^*$ being the magnitude of only imaginary eigenvalue(which is in pair) of $\bb{W}^*$. This gives us an idea of decomposing $\bb{m}_c$ into $\bb{m}_c=\ovl{\bb{m}}_c+\widetilde{\bb{m}}_c$ where $\ovl{\bb{m}}_c=\mathrm{Proj}_{ S}\bb{m}_c$ (so that $\norm{\widetilde{\bb{m}}_c}=\text{Dist}(\bb{m}_c, S)$, and $\widetilde{\bb{m}}_c\perp S)$, then RHS of Eq. (\ref{retrieval_soln}) is decomposed into
\begin{equation}\label{xr_integral_eq}
\begin{split}
\bb{x}_r(t) & = e^{-t}e^{t\bb{W}^*}\bb{x}_r(0) + \int_{0}^{t} e^{-s}\sin(\omega(t-s))\widetilde{\bb{m}}_c \,\dd s \\
& \quad\quad \quad\quad + \int_{0}^{t} e^{-s}\sbr{\cos(\lambda^*s)\sin(\omega(t-s))\ovl{\bb{m}}_c + \sin(\lambda^*s)\sin(\omega(t-s))\ovl{\bb{m}}_{c\wedge}}\,\dd s,
\end{split}
\end{equation} 
where $\ovl{\bb{m}}_{c\wedge}$ is decided by $\ovl{\bb{m}}_c$ in the means of relationship between $\bb{x}_{0\wedge}$ and $\bb{x}_0$ in Eq. (\ref{antisymm_flow}). 

Computing the asymptotic behaviour of each integral as $t\to\infty$ and using the fact that the term $e^{-t}e^{t\bb{W}^*}\bb{x}_r(0)$ decays to $\bb{0}$ as $t\to\infty$, then Eq. (\ref{xr_integral_eq}) turns out to be asymptotically approaching the following periodic function
\begin{equation}\label{retrieval_sol}
\boxed{\;
\bb{x}_r^*(t) = \frac{1}{\sqrt{\omega^2+1}}\sin\sbr{\omega t-\theta}\widetilde{\bb{m}}_c + c_{3,[\lambda^*,\omega]}(t)\ovl{\bb{m}}_c + c_{4,[\lambda^*,\omega]}(t)\ovl{\bb{m}}_{c\wedge},\;}
\end{equation}
where $\theta=\tan^{-1}\omega$, and periodic functions $c_{3,[\lambda^*,\omega]}(t)$, $c_{4,[\lambda^*,\omega]}(t)$ are from Lemma \ref{ci_lemma} with parameters $\lambda^*$ and $\omega$. 

This $\bb{x}_r^*(t)$ is also a solution of Eq. (\ref{system_retrieval}). To show this, directly substituting Eq. (\ref{retrieval_sol}) into Eq. (\ref{system_retrieval}) and simplifying by collecting the terms of each $\widetilde{\bb{m}}_c$, $\ovl{\bb{m}}_c$, and $\ovl{\bb{m}}_{c\wedge}$, then one can verify that
%\begin{align}\label{retrieval_cf}
%& \frac{\omega}{\sqrt{\omega^2+1}}\cos(\omega t-\theta)\widetilde{\bb{m}}_c+\omega c_{1,[\lambda^*,\omega]}(t)\ovl{\bb{m}}_c+\omega c_{2,[\lambda^*,\omega]}(t)\ovl{\bb{m}}_{c\wedge}\notag \\
%& \quad =-\frac{1}{\sqrt{\omega^2+1}}\sin(\omega t-\theta)\widetilde{\bb{m}}_c-c_{3,[\lambda^*,\omega]}(t)\ovl{\bb{m}}_c-c_{4,[\lambda^*,\omega]}(t)\ovl{\bb{m}}_{c\wedge} \notag\\
%& \quad\quad+ \frac{\lambda^*}{\nu^2}(\twist{\ovl{\bb{m}}_c}{\ovl{\bb{m}}_{c\wedge}})\sbr{\frac{1}{\sqrt{\omega^2+1}}\sin(\omega t+\theta)\widetilde{\bb{m}}_c+c_{3,[\lambda^*,\omega]}(t)\ovl{\bb{m}}_c+c_{4,[\lambda^*,\omega]}(t)\ovl{\bb{m}}_{c\wedge}} \notag \\
%& \quad\quad + \sin\omega t\bb{m}_c,
%\end{align}
\begin{align}
& \sbr{\frac{\omega}{\sqrt{\omega^2+1}}\cos(\omega t-\theta) +\frac{1}{\sqrt{\omega^2+1}}\sin(\omega t-\theta) }\widetilde{\bb{m}}_c \notag \\
&\quad\quad\quad\quad +(c_{1,[\lambda^*,\omega]}(t)+\omega c_{3,[\lambda^*,\omega]}(t)+\lambda^* c_{4,[\lambda^*,\omega]}(t))\ovl{\bb{m}}_c\notag \\
& \quad\quad\quad\quad +(c_{2,[\lambda^*,\omega]}(t)+\omega c_{4,[\lambda^*,\omega]}(t)-\lambda^* c_{3,[\lambda^*,\omega]}(t))\ovl{\bb{m}}_{c\wedge} - \sin\omega t\bb{m}_c =\bb{0},
\end{align}
where each $c_{1,[\lambda_*,\omega]}(t)$, $c_{2,[\lambda_*,\omega]}(t)$ arises from the differentiation of $c_{3,[\lambda_*,\omega]}(t)$ and $c_{4,[\lambda_*,\omega]}(t)$ with respect to $t$ as in Lemma \ref{ci_lemma}-1, and the alternate representation of $\bb{W}^*\in\wedge^2(S)$ with $\bb{W}^* = \alpha(\bb{v}\bb{u}^\top-\bb{u}\bb{v}^\top) = \frac{\lambda^*}{\nu^2}(\twist{\ovl{\bb{m}}_c}{\ovl{\bb{m}}_{c\wedge}})$ where $\nu = \norm{\ovl{\bb{m}}_c}$ has been used.

If one can show the equivalence of the LHS with $\bb{0}$, then it is done. In the LHS, the coefficient of $\widetilde{\bb{m}}_c$ is directly $\sin\omega t$, and the coefficient of $\ovl{\bb{m}}_c$ is also equivalent to $\sin\omega t$ by Lemma \ref{ci_lemma}-\textit{2b}. On the other hand, the coefficient of $\ovl{\bb{m}}_{c\wedge}$ is $0$ by Lemma \ref{ci_lemma}-\textit{2d}. Thus the LHS is actually $\sin\omega t(\widetilde{\bb{m}}_c+\ovl{\bb{m}}_c)-\sin\omega t \bb{m}_c$, which simply $\bb{0}$. This proves that the converging limit-cycle orbit $\bb{x}_r^*(t)$ of $\bb{x}_r(t)$ is also a solution of Eq. (\ref{system_retrieval}).

To show the remaining statements, let $\mathcal{M}$ be the retrievable subspace with respect to a representation set $\{\bb{m}_i \}_{i=1}^n$, and consider the case that $\bb{m}_c\not\in S_\perp$. Then $\ovl{\bb{m}}_c,\ovl{\bb{m}}_{c\wedge}\not=\bb{0}$, and since $\ovl{\bb{m}}_c,\ovl{\bb{m}}_{c\wedge}\in S =\spn\{\bb{u},\bb{v}\}$, one can directly see that also $\ovl{\bb{m}}_c,\ovl{\bb{m}}_{c\wedge}\in\mathcal{M}\setminus\{\bb{0}\}$. Now, we claim that the term $c_{3,[\lambda_*,\omega]}(t)\ovl{\bb{m}}_c + c_{4,[\lambda_*,\omega]}(t)\ovl{\bb{m}}_{c\wedge}$ in Eq. (\ref{retrieval_sol}) always lies in $\mathcal{M}\setminus\{\bb{0}\}$. This can be shown from the following alternate expressions of $c_{3,[\lambda^*,\omega]}(t)$ and $c_{4,[\lambda^*,\omega]}(t)$:
\begin{align}\label{c3c4alt}
& c_{3,[\lambda^*,\omega]}(t) = \sqrt{\alpha^2+\beta^2}\sin(\omega t+\Delta_1) \quad\mathrm{and}\quad c_{4,[\lambda^*,\omega]}(t) = \sqrt{\alpha^2+\beta^2}\sin(\omega t+\Delta_2), \\
& \;\;\;\where\quad\begin{dcases}
\alpha = \frac{1}{2}\sbr{\frac{\cos\theta_{-,[\lambda^*,\omega]}}{\sqrt{\Phi_{-,[\lambda^*,\omega]}}} +\frac{\cos\theta_{+,[\lambda^*,\omega]}}{\sqrt{\Phi_{+,[\lambda^*,\omega]}}} }  \\
\beta = \frac{1}{2}\sbr{\frac{\sin\theta_{-,[\lambda^*,\omega]}}{\sqrt{\Phi_{-,[\lambda^*,\omega]}}} +\frac{\sin\theta_{+,[\lambda^*,\omega]}}{\sqrt{\Phi_{+,[\lambda^*,\omega]}}} }
\end{dcases} \quad\mathrm{and}\quad\begin{dcases}
\Delta_1 = \tan^{-1}\sbr{-\frac{\beta}{\alpha}}\\
\Delta_2 = \tan^{-1}\sbr{\frac{\alpha}{\beta}}.
\end{dcases}
\end{align}
From Eq. (\ref{c3c4alt}), one can read that the only condition making $c_{3,[\lambda^*,\omega]}$ and $c_{4,[\lambda^*,\omega]}$ to vanish simultaneously is $\Delta_1 = \Delta_2$, and the bijective property of the arctangent function implies
\begin{align}
-\frac{\beta}{\alpha}=\frac{\alpha}{\beta}\quad \Longleftrightarrow \quad \alpha^2+\beta^2 = 0,
\end{align}
thus yielding $\alpha=\beta=0$, which only is a pointless triviality.

From this, one can assure that $\bb{x}_r^*(t)$ must belongs to $\mathcal{M}\setminus\{\bb{0}\}$ on instances that making the coefficient of $\widetilde{\bb{m}}_c$ in Eq. (\ref{retrieval_sol}) to vanish. Denoting such time $t$ as $t=t^\dagger$, we derive that such $t^\dagger$ must satisfy $\omega t^\dagger-\theta = n\pi$, $n\in\ZZ$, that is,
\begin{equation}
\boxed{\;
t^\dagger = \frac{1}{\omega}\tan^{-1}\omega + n\frac{\pi}{\omega},\quad n\in\ZZ,\;(t^\dagger >0),\;}
\end{equation}
which yields Eq. (\ref{t_dagger}) indicating periodic retrieval. This proves that $\bb{x}^*_r(t)$ with $t=t^\dagger$ is always retrievable unless $\bb{m}_c\not\in S_\perp$. Besides, if $\bb{m}_c\in\mathcal{M}$, then one can easily see $\widetilde{\bb{m}}_c\in\mathcal{M}$ thus $\bb{x}_r^*(t)\in\mathcal{M}\setminus\{\bb{0}\}$ for all $t$, so always being retrievable. 

On the other hand, considering the case when $\bb{m}_c\in S_\perp$, first suppose that also $\bb{m}_c\in\mathcal{M}$. Then, $\ovl{\bb{m}}_c$, $\ovl{\bb{m}}_{c\wedge}=\bb{0}$, but $\widetilde{\bb{m}}_c\not=0$. Thus $\bb{x}^*_r(t)\in\mathcal{M}$ for all $t$, but especially only on $t=t^\dagger$, $\bb{x}^*_r(t)=\bb{0}$. In contrary, if $\bb{m}_c\not\in\mathcal{M}$, then also $\ovl{\bb{m}}_c$, $\ovl{\bb{m}}_{c\wedge}=\bb{0}$, and even $\widetilde{\bb{m}}_c\not\in\mathcal{M}$, therefore $\bb{x}^*_r(t)\not\in\mathcal{M}\setminus\{\bb{0}\}$ for all $t$, so never becoming retrievable.

Summing up above results, the retrievability conditions in main text page 6 and Theorem 2 have been proved.
%such non-trivial $\bb{m}_c$ exists when $N>n$ by the following:
%\begin{customlem}{C}\label{s_cerp_m}
%If $N=n$, then $ S_\perp\setminus\mathcal{M}=\phi$. Otherwise, if $N>n$, then $ S_\perp\setminus\mathcal{M}\not=\phi$.
%\begin{proof}
%When $N=n$, it is trivial since $\mathcal{M}=\RR^N$. When $N>n$, there must exist directions $\bb{h}_i$, ($i=1,\dots,N-n$) such that $\bb{h}_i\perp\mathcal{M}$, $\forall i$. Now, consider any $\sum_i\alpha_i\bb{h}_i\in\mathcal{M}_\perp\setminus\{\bb{0}\}$ so that $\sum_i\alpha_i\bb{h}_i\not\in\mathcal{M}$. Since $ S\subset\mathcal{M}$, so $\mathcal{M}_\perp\subset S_\perp$, therefore $\sum_{i}\alpha_i\bb{h}_i\in\bb{S}_\perp$. This proves $ S_\perp\setminus\mathcal{M}\not=\phi$ whenever $N>n$.
%\end{proof}
%\end{customlem}
%$\widetilde{\bb{m}}_c=\bb{m}_c\not\in\mathcal{M}\setminus\{\bb{0}\}$, which directly leads to $\bb{x}_r^*(t)\not\in\mathcal{M}\setminus\{\bb{0}\}$, $\forall t>0$, especially with $\bb{x}_r^*(t^\dagger)=\bb{0}$. This proves statement \textit{3}. 
%
%Finally, if $\bb{m}_c\in S_\perp$ and $\bb{m}_c\in\mathcal{M}$, then we directly see $\ovl{\bb{m}}_c,\ovl{\bb{m}}_{c\wedge}=\bb{0}$ and $\widetilde{\bb{m}}_c\in\mathcal{M}$, thus $\forall t>0$ except $t = \frac{1}{\omega}\tan^{-1}\omega\mod \frac{\pi}{\omega}$, $\bb{x}_r^*(t)\in\mathcal{M}$, since such $t$ leads $\bb{x}_r^*(t) = \bb{0}$, thus proving the statement \textit{4}.
\hfill $\blacksquare$

\subsubsection*{\textit{B4: Proof of Theorem 3}}

One can directly use $\cos\theta_i=\frac{\norm{\mathrm{Proj}_{S(\xi_1,\dots,\xi_n)}\bb{m}_i}}{\norm{\bb{m}_i}}$ where $S(\xi_1,\dots,\xi_n)$ denotes the memory plane determined with the choice of $\{\xi_i\}_{i=1}^n$. Firstly, one can generally observe that for any $\bb{m}\in\RR^N$ and $S=\spn\{\bb{u},\bb{v}\}$,
\begin{align}
\mathrm{Proj}_{S}\bb{m} = \mathrm{Proj}_{\spn\{\bb{u},\bb{v}\}}\bb{m} = \frac{\bb{m}^\top\bb{u}}{\norm{\bb{u}}}\bb{u} + \frac{\bb{m}^\top\bb{u}_\perp}{\norm{\bb{u}_\perp}}\bb{u}_\perp,
\end{align}
where $\bb{u}_\perp\in \spn\{\bb{u},\bb{v}\}$ satisfying $\bb{u}_\perp\perp\bb{u}$ and $\norm{\bb{u}_\perp}=\norm{\bb{u}}$, in which can be specifically expressed as in Eq. (\ref{thm_perpexpression}). Thus substituting it into above equation yields
\begin{align}
\mathrm{Proj}_{S}\bb{m}_i = \frac{1}{1-\mu^2}\sbr{\frac{\bb{m}_i^\top\bb{u}}{\eta_1^2}-\mu\frac{\bb{m}_i^\top\bb{v}}{\eta_1\eta_2}}\bb{u} + \frac{1}{1-\mu^2}\sbr{\frac{\bb{m}_i^\top\bb{v}}{\eta_2^2}-\mu\frac{\bb{m}_i^\top\bb{u}}{\eta_1\eta_2}}\bb{v},
\end{align}
where $\eta_1=\norm{\bb{u}}$, $\eta_2=\norm{\bb{v}}$ and $\mu = \bb{u}^\top\bb{v}/\sbr{\norm{\bb{u}}\norm{\bb{v}}}$. Besides, from the fact
\begin{align}
\norm{\mathrm{Proj}_S\bb{m}_i}^2 = \bb{m}_i\cdot\mathrm{Proj}_S\bb{m}_i,
\end{align} 
for $\bb{u},\bb{v}$ defined as in Lemma A, one can directly read that
\begin{align}\label{proj_s}
\norm{\mathrm{Proj}_{S(\xi_1,\dots,\xi_n)}\bb{m}_i}=\sqrt{\frac{\sbr{\frac{\bb{m}_i^\top\bb{u}}{\eta_1}}^2 + \sbr{\frac{\bb{m}_i^\top\bb{v}}{\eta_2}}^2 -2\mu\frac{\sbr{\bb{m}_i^\top\bb{u}}\sbr{\bb{m}_i^\top\bb{v}}}{\eta_1\eta_2}}{1-\mu^2}}. 
\end{align}

Let $\norm{\bb{m}_i}=l$. Since $\bb{u}=-\boldsymbol\Psi\sin\boldsymbol\xi = \sum_{i=1}^n \sin\xi_i\bb{m}_i$, and $\bb{v}=\boldsymbol\Psi\cos\boldsymbol\xi = \sum_{i=1}^n \cos\xi_i\bb{m}_i$, the orthogonality of $\{\bb{m}_i\}_{i=1}^n$ guarantees $\bb{m}_i^\top\bb{u} = -\sin\xi_i$, and $\bb{m}_i^\top\bb{v} = \cos\xi_i$. Further, one can easily verify that
\begin{align*}
\eta_1 = l^2\sum_{j=1}^n\sin^2\xi_j,\quad \eta_2 = l^2\sum_{j=1}^n\cos^2\xi_j\quad \mathrm{and}\; \mu = -\frac{\sum_{j=1}^{n}\sin\xi_j\cos\xi_j}{\sqrt{\sbr{\sum_{j=1}^n\sin^2\xi_j}\sbr{\sum_{j=1}^n\cos^2\xi_j}}},
\end{align*}
so substituting these expressions into Eq. (\ref{proj_s}) and completing tedious simplification procedure, we finally deduce
\begin{equation}\label{uv_xi2}
\frac{\norm{\mathrm{Proj}_{S(\xi_1,\dots,\xi_n)}\bb{m}_i}}{\norm{\bb{m}_i}} = \sqrt{\frac{\sum\limits_{j=1}^n \sin^2(\xi_j-\xi_i)}{\sum\limits_{\substack{j,k=1\\j>k}}^{n}\sin^2(\xi_j-\xi_k)}}.
\end{equation}

Note that this value does not depend on $l=\norm{\bb{m}}_i$. Now, consider the following double summation $\sum_{i,j=1}^{n}\sin^2(\xi_j-\xi_i)$. This is exactly the sum with respect to $i$ performed to the squared numerator of the last term in Eq. (\ref{uv_xi2}). Moreover, $\sin^2(\xi_j-\xi_i)=\sin^2(\xi_i-\xi_j)$ and is zero when $j=i$, thus we read that \begin{equation}
\sum_{i,j=1}^{n}\sin^2(\xi_j-\xi_i)=2\sum_{\substack{i,j=1 \\ j>i}}^n\sin^2(\xi_j-\xi_i),
\end{equation} 
which the term $\sum_{\substack{i,j=1 \\ j>i}}^n\sin^2(\xi_j-\xi_i)$ is identical the squared denominator of the last term in Eq. (\ref{uv_xi2}). This directly leads to the following strong result:
\begin{equation}\label{xi_sum}
\boxed{\;
\sum_{i=1}^{n}\sbr{\frac{\norm{\mathrm{Proj}_{S(\xi_1,\dots,\xi_n)}\bb{m}_i}}{\norm{\bb{m}_i}}}^2 = 2.\;}
\end{equation}
From this, we see that by the Cauchy-Schwarz inequality, the maximum of $\langle\cos\theta_i\rangle_i=\linebreak\frac{1}{n}\sum_{i=1}^{n}\frac{\norm{\mathrm{Proj}_{S(\xi_1,\dots,\xi_n)}\bb{m}_i}}{\norm{\bb{m}_i}}$ is achieved with value $\sqrt{\frac{2}{n}}$ when each $\cos\theta_i=\frac{\norm{\mathrm{Proj}_{S(\xi_1,\dots,\xi_n)}\bb{m}_i}}{\norm{\bb{m}_i}}=\frac{\sqrt{2n}}{n}=\sqrt{\frac{2}{n}}$ for all $i=1,\dots,n$, so proving Eq. (\ref{max_perf}). 

However, finding the possible distributions of $\xi_i$ achieving the maximum is quite difficult, but we claim that such distribution exists, and one family of those are given as in (\ref{max_perf}). To show this, first suppose that each $\xi_i$ is chosen as (\ref{max_perf}) but with zero shifts, i.e., $\alpha=0$, and denote such values with $\bar{\xi}_i$. We first verify that $\bb{u}\perp\bb{v}$ in this case. Observe that when $n$ is even,
\begin{align}
\bb{u}^\top\bb{v} &= -\sum_{i=1}^{n}\sin\bar{\xi}_i\cos\bar{\xi}_i  = -\sin\bar{\xi}_1\cos\bar{\xi}_1-\sum_{i=2}^{n}\sin\bar{\xi}_i\cos\bar{\xi}_i \notag \\ 
& = 0-\sum_{i=2}^{n/2}\sbr{\sin\bar{\xi}_i\cos\bar{\xi}_i+\sin\bar{\xi}_{n-i+2}\cos\bar{\xi}_{n-i+2}} + \sin\bar{\xi}_{n/2+1}\cos\bar{\xi}_{n/2+1} \notag \\
& = -\sum_{i=2}^{n/2}\sbr{\sin\bar{\xi}_i\cos\bar{\xi}_i + \sin(\pi-\bar{\xi}_i)\cos(\pi-\bar{\xi}_i)}+\sin\frac{\pi}{2}\cos\frac{\pi}{2} = 0,
\end{align}
and similarly when $n$ is odd,
\begin{align}
\bb{u}^\top\bb{v} & =-\sum_{i=1}^{n}\sin\bar{\xi}_i\cos\bar{\xi}_i \notag \\
& = -\sin\bar{\xi}_1\cos\bar{\xi}_1 -\sum_{i=2}^{(n+1)/2}\sbr{\sin\bar{\xi}_i\cos\bar{\xi}_i + \sin\bar{\xi}_{n-i+2}\cos\bar{\xi}_{n-i+2}} \notag \\
& = 0 -\sum_{i=2}^{(n+1)/2}\sbr{\sin\bar{\xi}_i\cos\bar{\xi}_i + \sin(\pi-\bar{\xi}_i)\cos(\pi-\bar{\xi}_i)} = 0.
\end{align}
Therefore, $\bb{u}\perp\bb{v}$, so simply considering a $\mu = 0$ case in Eq. (\ref{proj_s}), we have
\begin{equation}\label{temp_theta_i}
\begin{split}
\frac{\norm{\mathrm{Proj}_{S(\bar{\xi}_1,\dots,\bar{\xi}_n)}\bb{m}_i}}{\norm{\bb{m}_i}}= \sqrt{\frac{\sin^2\bar{\xi}_i}{\sum_{j=1}^n\sin^2\bar{\xi}_j} + \frac{\cos^2\bar{\xi}_i}{\sum_{j=1}^n\cos^2\bar{\xi}_j} },\quad i=1,\dots,n.
\end{split}
\end{equation}

Here, one can even show that
\begin{equation}\label{sum_xi}
\sum_{j=1}^n\sin^2\bar{\xi}_j = \sum_{j=1}^n\cos^2\bar{\xi}_j = \frac{n}{2}
\end{equation} 
by observing the following: From Riemann integral,
\begin{equation}\label{Riemann_approx}
\begin{split}
\Delta\sum_{j=1}^n\sin^2\bar{\xi}_j \approx\int_{0}^{\pi}\sin^2\theta\;\dd\theta  =  \frac{\pi}{2},\quad\;\;\Delta\sum_{j=1}^n\cos^2\bar{\xi}_j \approx\int_{0}^{\pi}\cos^2\theta\;\dd\theta = \frac{\pi}{2}
\end{split}
\end{equation}
as $n\to\infty$ where $\Delta=\pi/n$ being the interval between each sampling points $\bar{\xi}_j$. However, by the symmetry of functions $\cos^2\theta$ and $\sin^2\theta$ on interval $[0,\pi]$ and the arithmetically sequenced property of $\bar{\xi}_j$, one can luckily confirm that the approximation (\ref{Riemann_approx}) is actually an equality for all $n$. Thus we finally have 
\begin{equation}
\begin{split}
\sum_{j=1}^{n}\sin^2\bar{\xi}_j = \sum_{j=1}^{n}\cos^2\bar{\xi}_j & = \frac{\pi}{2\Delta} = \frac{\pi}{2\cdot \frac{\pi}{n}},
\end{split}
\end{equation}
which yields Eq. (\ref{sum_xi}). Therefore, we can now write Eq. (\ref{temp_theta_i}) simply as
\begin{equation}
\begin{split}
\frac{\norm{\mathrm{Proj}_{S(\bar{\xi}_1,\dots,\bar{\xi}_n)}\bb{m}_i}}{\norm{\bb{m}_i}} & = \sqrt{\frac{2(\sin^2\bar{\xi}_i+\cos^2\bar{\xi}_i)}{n}} = \sqrt{\frac{2}{n}}, \quad i=1,\dots,n.
\end{split}
\end{equation}
This indicates that the value of $\norm{\mathrm{Proj}_{S(\bar{\xi}_1,\dots,\bar{\xi}_n)}\bb{m}_i}$ is constant throughout every $i=1,\dots,n$ with value $\sqrt{2/n}$, so such set of $\{\bar{\xi}_i\}_{i=1}^n$ (i.e., in Eq. (\ref{max_perf}) with $\alpha = 0$) can achieve \linebreak $\max_{\xi_1,\dots,\xi_n}\frac{\norm{\mathrm{Proj}_{S(\xi_1,\dots,\xi_n)}\bb{m}_i}}{\norm{\bb{m}_i}} =\sqrt{2n}$.

Lastly, for the remaining $\alpha\not=0$ case, i.e., $0<\alpha<\frac{\pi}{n}$, recall that $\xi^*_i  = \bar{\xi}_i+\alpha$. Let's denote $\bb{b}_{(\xi_1,\dots,\xi_n)}(t) = \sum_{i=1}^n\sin(\omega t-\xi_i)\bb{m}_i$ as the input orbit generated by $\{\xi_i\}_{i=1}^n$. Then, one can easily see that $\bb{b}_{(\xi^*_1,\dots,\xi^*_n)}(t) = \bb{b}_{(\bar{\xi}_1,\dots,\bar{\xi}_n)}\sbr{t+\frac{\alpha}{\omega}}$ for any $t$, so the orbit of $\bb{b}_{(\xi^*_1,\dots,\xi^*_n)}$ and $\bb{b}_{(\bar{\xi}_1,\dots,\bar{\xi}_n)}$ is actually identical thus sharing the same plane, i.e., $S(\xi^*_1,\dots,\xi^*_n)\equiv S(\bar{\xi}_1,\dots,\bar{\xi}_n)$ from Lemma A. Thus, one must have $\langle\cos\theta_i(\xi^*_1,\dots,\xi^*_n)\rangle_i= \langle\cos\theta_i(\bar{\xi}_1,\dots,\bar{\xi}_n)\rangle_i=\sqrt{\frac{2}{n}}$, which implies that $\xi^*_i = \bar{\xi}_i+\alpha=\frac{\pi}{n}(i-1)+\alpha$ also achieves the maximum of $\langle\cos\theta_i\rangle_i$.
\hfill$\blacksquare$
\bigskip
\subsection*{Appendix C: Stability Analysis of the Periodic Solution $(\bb{x}^*(t),\bb{W}^*)$} 

\subsubsection*{\textit{C1: Derivation of the Variational Equation, Eq. (\ref{variational_dde})}}

First, rewriting the original system (\ref{system}) into a general form, then
\begin{equation}\label{gen_sys}
\begin{dcases}
\dot{\bb{x}} = \bb{f}(\bb{x},\bb{W}) \\
\dot{\bb{W}} = \bb{G}(\bb{x},\bb{x}_\tau,\bb{W}) \\
\end{dcases}\;\text{where}\;
\left.\begin{array}{l}
\bb{f}(\bb{x},\bb{W}) = -\bb{x}+\bb{W}\bb{x}+ \bb{b}(t),\\
\bb{G}(\bb{x},\bb{x}_\tau,\bb{W}) = -\gamma \bb{W} + \rho\sbr{\bb{x}\bb{x}_\tau^\top-\bb{x}_\tau\bb{x}^\top}.
\end{array}\right.
\end{equation}
Considering deviation $\bb{x}(t) = \bb{x}^*(t) + \delta\bb{x}(t)$ and $\bb{W}(t) = \bb{W}^*+\delta \bb{W}(t)$ from reference trajectory $(\bb{x}^*,\bb{W}^*)$, we have
\begin{equation}
\begin{dcases}
\dot{\bb{x}}^*+\dot{\delta\bb{x}} = \bb{f}(\bb{x}^*+\delta\bb{x},\bb{W}^*+\delta\bb{W})\\
\dot{\bb{W}}^*+\dot{\delta\bb{W}} = \bb{G}(\bb{x}^*+\delta\bb{x},\bb{x}^*_\tau+\delta\bb{x}_\tau,\bb{W}^*+\delta\bb{W}).
\end{dcases}
\end{equation}
 
Now, applying first-ordered Taylor expansion on $(\bb{x}^*,\bb{W}^*)$ to each RHS and using $\dot{\bb{x}}^*=\bb{f}(\bb{x}^*,\bb{W}^*)$ and $\dot{\bb{W}}^* = \bb{G}(\bb{x}^*,\bb{x}^*_\tau,\bb{W}^*)=\bb{O}$, we get
\begin{equation}
\begin{dcases}
\dot{\delta\bb{x}} = \at{\frac{\pd \bb{f}}{\pd\bb{x}}}_{(\bb{x}^*,\bb{W}^*)}\cdot\delta\bb{x} + \at{\frac{\pd \bb{f}}{\pd \bb{W}}}_{(\bb{x}^*,\bb{W}^*)}:\delta \bb{W} \\
\dot{\delta \bb{W}} = \at{\frac{\pd \bb{G}}{\pd\bb{x}}}_{(\bb{x}^*,\bb{x}^*_\tau,\bb{W}^*)}\cdot\delta\bb{x}+\at{\frac{\pd \bb{G}}{\pd\bb{x}_\tau}}_{(\bb{x}^*,\bb{x}^*_\tau,\bb{W}^*)}\cdot\delta\bb{x}_\tau + \at{\frac{\pd \bb{G}}{\pd \bb{W}}}_{(\bb{x}^*,\bb{x}^*_\tau,\bb{W}^*)}:\delta \bb{W},
\end{dcases}
\end{equation}
where $:$ is used for the \textit{double dot product} notation. Now computing each tensor-represented Jacobians, firstly we immediately see $\frac{\pd\bb{f}}{\pd\bb{x}} = -\bb{I}+\bb{W}$, therefore
\begin{equation}
\at{\frac{\pd\bb{f}}{\pd\bb{x}}}_{(\bb{x}^*,\bb{W}^*)}\cdot\delta\bb{x} = (-\bb{I}+\bb{W}^*)\delta\bb{x}.
\end{equation}
For the remaining ones, observe that $\frac{\pd\bb{f}}{\pd\bb{W}}$ is a third-order tensor and each element can be found by\smallskip
\begin{equation}
\begin{split}
\sbr{\frac{\pd\bb{f}}{\pd\bb{W}}}_{ijk} & = \frac{\pd f_i}{\pd W_{jk}} = \frac{\pd}{\pd W_{jk}}\sbr{-x_i+\sum_l W_{il}x_l+b_i(t)}\\
& = \delta_{ij}\delta_{kl}x_l = \delta_{ij}x_k.
\end{split}
\end{equation}
Thus if write $\bb{e}^i$ as the $i$-th coordinate Euclidean canonical vector (i.e., $(\bb{e}^i)_j=\delta_{ij}$), then one can have
\begin{equation} 
\begin{split}
\at{\frac{\pd\bb{f}}{\pd\bb{W}}}_{(\bb{x}^*,\bb{W}^*)}:\delta\bb{W} & = \sum_{i,j,k}\delta_{ij}x_k^*(\bb{e}^i\otimes\bb{e}^j\otimes\bb{e}^k):\sum_{l,m}\delta W_{lm}(\bb{e}^l\otimes\bb{e}^m) \\
& = \sum_{i,j,k,l,m}\delta_{ij}x_k^*\delta_{jl}\delta_{km}\delta W_{lm}\,\bb{e}^i  = \sum_{i,j,k}\delta_{ij}x_k^*\delta W_{jk}\,\bb{e}^i \\
&= \sum_{i,k}\delta W_{ik}x_k^*\,\bb{e}^i = \delta\bb{W}\bb{x}^*.
\end{split}
\end{equation}

By similar computations, for remaining terms one can easily verify that $\at{\frac{\pd \bb{g}}{\pd\bb{x}}}_{(\bb{x}^*,\bb{x}^*_\tau,\bb{W}^*)}\cdot\delta\bb{x}= \rho\sbr{\delta\bb{x}\,\bb{x}_\tau^{*\top}-\bb{x}_\tau^*\delta\bb{x}^\top}$, $\at{\frac{\pd \bb{g}}{\pd\bb{x}_\tau}}_{(\bb{x}^*,\bb{x}^*_\tau,\bb{W}^*)}\cdot\delta\bb{x}_\tau = \rho\sbr{\bb{x}^*\,\delta\bb{x}_\tau^\top-\delta\bb{x}_\tau\bb{x}^{*\top}}$, and $\at{\frac{\pd \bb{g}}{\pd \bb{W}}}_{(\bb{x}^*,\bb{x}^*_\tau,\bb{W}^*)}:\delta \bb{W} =-\gamma\,\delta\bb{W}$.
Therefore summing up the results, we finally get
\begin{equation}
\boxed{\;
\begin{dcases}
\dot{\delta\bb{x}} = (-\bb{I}+\bb{W}^*)\,\delta\bb{x} + \delta \bb{W}\bb{x}^* \\
\dot{\delta \bb{W}} = -\gamma\,\delta \bb{W} + \rho\sbr{\delta\bb{x}\,\bb{x}_\tau^{*\top}-\bb{x}_\tau^*\,\delta\bb{x}^\top + \bb{x}^*\,\delta\bb{x}_\tau^\top - \delta\bb{x}_\tau\,\bb{x}^{*\top}},
\end{dcases}\;}
\end{equation}
and this is the variational equation, Eq. (\ref{variational_dde}).
\hfill$\blacksquare$

\subsubsection*{\textit{C2: Computational Method for Estimating Maximal Lyapunov Exponent}}

The method of computation directly follows \cite{farmer1982chaotic}. First, the DDE (\ref{variational_dde}), say, $\dot{\bb{U}} = \bb{F}(\bb{U},\bb{U}_\tau)$, where $\bb{U}\in\RR^{N+N^2}$ represents the collection of all components of $\delta\bb{x}$ and $\delta\bb{W}$, can be approximated with some conjugate discrete finite dimensional map
\begin{align}
\bar{\bb{F}}:\underbrace{\RR^{N+N^2}\times\cdots\times\RR^{N+N^2}}_{d}\to \underbrace{\RR^{N+N^2}\times\cdots\times\RR^{N+N^2}}_d,
\end{align}
having variables $\bar{\bb{U}}^n\in\RR^{N+N^2}$, $n=1,\dots,d$, which
\begin{equation}\label{ddemap}
(\bar{\bb{U}}^1,\cdots,\bar{\bb{U}}^{d-1},\bar{\bb{U}}^d) = (\bb{U}(t-(d-1)\Delta t),\cdots,\bb{U}(t-\Delta t),\bb{U}(t)),\quad\sbr{\Delta t=\frac{\tau}{d-1}},
\end{equation}
so that the each iteration $\bar{\bb{U}}(k+1) =\bar{\bb{F}}(\bar{\bb{U}}(k))$ for $\bar{\bb{U}}$ represents the mapping of $\bar{\bb{U}}=(\bar{\bb{U}}^1,\dots,\bar{\bb{U}}^d)$ on time $t$ to $t+\tau+\Delta t$. 
%For example, if $\phi_i$ is an initial function of Eq. (\ref{variational_dde}) for $\bb{x}_i$ on interval $[-\tau,0]$, then 
%\begin{equation}
%\begin{split}
%\bar{\bb{x}}_i(0) & = \begin{bmatrix}
%\bar{x}_i^1(0)& \bar{x}_i^2(0) &\cdots & \bar{x}_i^{d-1}(0) & \bar{x}_i^d(0)
%\end{bmatrix} \\
%& = \begin{bmatrix}
% \phi_i(-\tau) & \phi_i\sbr{-\frac{d-2}{d-1}\tau} & \cdots &\phi_i\sbr{-\frac{1}{d-1}\tau} & \phi_i(0)
%\end{bmatrix}.
%\end{split}
%\end{equation}
As the initial choice of $\bar{\bb{U}}$ is given by sampled discrete points on $t\in[-\tau,0]$, this map starts to generate the approximated solution on interval $[\Delta t,\;\tau+\Delta t]$, $[\tau+2\Delta t,\;2\tau+2\Delta t]$ and so on. 

The discrete map $\bar{\bb{F}}$ conjugate to $\bb{F}$ can be found by any convenient integration techniques. Simply, for example, Euler-method integration takes
\begin{equation}\label{lyap_euler}
\begin{split}
\bar{\bb{U}}^1(k+1) & = \bar{\bb{U}}^d(k) + \bb{F}(\bar{\bb{U}}^d(k),\bar{\bb{U}}^1(k))\Delta t, \\
\text{and\;for}\quad 1<i\le d;\quad\bar{\bb{U}}^i(k+1) & = \bar{\bb{U}}^{i-1}(k+1) + \bb{F}(\bar{\bb{U}}^{i-1}(k+1),\bar{\bb{U}}^{i}(k))\Delta t.
\end{split}
\end{equation}

%Now, exactly same works can be done for Eq. (\ref{variational_dde}). Note that it is also a DDE, the initial condition representing deviation for each component of $\bb{x}$ and $\bb{W}$ must be a function on interval $[-\tau,0]$. 

Now, setting $\bar{\bb{U}}(0)$ containing all of the discrete-sampled initial data of each $\delta\bar{\bb{x}}_i$, $\delta\bar{\bb{W}}_{ij}\in\RR^{d}$ and obtaining the evolution of $\bar{\bb{U}}$ for each step, then the rate of exponential growth of universal deviation (the collection of every deviations)
\begin{equation}
[\bar{\bb{U}}](k) = \begin{bmatrix}
\delta\bar{\bb{x}}_1(k);\;\cdots\;;\delta\bar{\bb{x}}_N(k);\delta\bar{\bb{W}}_{11}(k);\;\cdots\;;\delta\bar{\bb{W}}_{NN}(k) 
\end{bmatrix}\in\RR^{d(N+N^2)}
\end{equation}
where `$;$' denotes the vertical concatenation, is estimated by directly computing the value
\begin{equation}\label{lyap_comp}
\boxed{\;
\lambda_{\text{max}}=\lim_{K\to\infty}\frac{1}{K(\tau+\Delta t)}\sum_{k=1}^{K}\ln\sbr{\frac{\norm{[\bar{\bb{U}}](k)}}{\norm{[\bar{\bb{U}}](k-1)}}}.\;}
\end{equation} 
This value $\lambda_{\mathrm{max}}$, turns out to be the maximal rate of exponential evolution of the universal deviation and in fact is the MLE, and its convergence as $K\to\infty$ is well known  \cite{farmer1982chaotic}. %Here, our reference trajectory $(\bb{x}^*,\bb{W}^*)$ is a theoretical periodic orbit that always yields trivial exponent (which is always 0), but $\lambda_{\text{max}}$ measured by (\ref{lyap_comp}) with general random initial deviations will be the maximal one except that trivial exponent.

\bigskip
\normalsize
\bibliography{memory_bib.bib}
\bibliographystyle{ieeetr}

\end{document}